\documentstyle[preprint,aps,eqsecnum,floats,epsfig,aps]{revtex}
\begin{document}
\def\bit{\begin{itemize}}
\def\eit{\end{itemize}}
\def\bnu{\begin{enumerate}}
\def\enu{\end{enumerate}}
\def\sss{\scriptscriptstyle}
\def\ss{\scriptstyle}
\def\rh {\hat{r}}
\def\fb {{\bf f}}
\def\Sb{ {\bf S}}
\def\xb {{\bf x}}
\def\Xb {{\bf X}}
\def\A {{{\cal A}}}
\def\B {{{\cal B}}}
\def\Q {{{\cal Q}}}
\def\R {{{\cal R}}}
\def\Qb{ {\bf Q}}
\def\nn{\nonumber }
\def\qb {{\bf q}}
\def\mbt{\mbox{\boldmath$\tau$}}
\def\mbr{\mbox{\boldmath$\rho$}}
\def\bin#1#2{\left(\negthinspace\begin{array}{c}#1\\#2\end{array}\right)}
\def\mbp{\mbox{\boldmath$\phi$}}
\def\mbxi{\mbox{\boldmath$\xi$}}
\def\mbpi{\mbox{\boldmath$\pi$}}
\def\boxp{\mbox{\boldmath$p$}}
\def\boxq{\mbox{\boldmath$q$}}
\def\mbs{\mbox{\boldmath$\sigma$}}
\def\mbn{\mbox{\boldmath$\nabla$}}
\def\H {{{\cal H}}}
\def\M {{{\cal M}}}
\def\x{\times}
\def\Ket#1{||#1 \rangle}
\def\Bra#1{\langle #1||}
\def\lsim{\:\raisebox{-0.5ex}{$\stackrel{\textstyle<}{\sim}$}\:}
\def\gsim{\:\raisebox{-0.5ex}{$\stackrel{\textstyle>}{\sim}$}\:}
\def\ie{{\em i.e., }}
\def\nn{\nonumber }
\def\be{\begin{equation}}
\def\ee{\end{equation}}
\def\br{\begin{eqnarray}}
\def\er{\end{eqnarray}}
\def\brn{\begin{eqnarray*}}
\def\ern{\end{eqnarray*}}
\def\etc{ {\it etc}}
\def\qb {{\bf q}}
\def\pb {{\bf p}}
\def\Pb{ {\bf P}}
\def\rb {{\bf r}}
\def\Rb{ {\bf R}}
\def\e {{\epsilon}}
\def\mbl{\mbox{\boldmath$\lambda$}}
\def\mbs{\mbox{\boldmath$\sigma$}}
\def\mbt{\mbox{\boldmath$\tau$}}
\def\bra#1{\langle #1|}
\def\ket#1{|#1 \rangle}
\def\rf#1{{(\ref{#1})}}
\def\ov#1#2{\langle #1 | #2  \rangle }
\def\sixj#1#2#3#4#5#6{\left\{\negthinspace\begin{array}{ccc}
#1&#2&#3\\#4&#5&#6\end{array}\right\}}
\def\ninj#1#2#3#4#5#6#7#8#9{\left\{\negthinspace\begin{array}{ccc}
#1&#2&#3\\#4&#5&#6\\#7&#8&#9\end{array}\right\}}
\def\ss{\scriptstyle}
\def\go{\rightarrow  }
\def\etal{{\it et al. }}
\def\E {{{\cal E}}}
\def\I {{{\cal I}}}
\def\V {{{\cal V}}}
\def\sqi{\frac{1}{\sqrt{2}}}
\def\fot{\frac{1}{2}}

\draft

\title{Hypernuclear Weak Decay Puzzle}

\author{C. Barbero$^{1,4}$, D. Horvat$^{2}$, F. Krmpoti\'c$^{1}$,
T. T. S. Kuo$^{5}$, Z. Naran\v ci\' c$^{2}$ and D. Tadi\'c$^{3}$}

\address{$^1$Departamento de F\'{\i}sica, Universidad Nacional de
La Plata, C. C. 67, 1900 La Plata, Argentina}
\address{$^2$Department of Physics, Faculty of Electrical
Engineering,\\ University of Zagreb, 10\,000 Zagreb, Croatia}
\address{$^3$Physics Department,University of Zagreb, 10\,000 Zagreb, Croatia}
\address{$^4$Centro Brasileiro de Pesquisas F\'{\i}sicas\\
Rua Dr. Xavier Sigaud 150, 22290-180 Rio de Janeiro-RJ, Brazil}
\address{$^5$Department of Physics, SUNY-Stony Brook, Stony Brook, New York 11794}

\date{\today}
\maketitle
\begin{abstract}
A general shell model formalism for the nonmesonic weak decay of the
hypernuclei has been developed.
It involves a partial wave expansion of the emitted nucleon waves,
preserves   naturally  the antisymmetrization between the
escaping particles and the residual  core, and contains as a particular case
the weak $\Lambda$-core coupling formalism.
The Extreme Particle-Hole Model and  the Quasiparticle Tamm-Dancoff Approximation
are  explicitly worked out. It is shown  that the  nuclear structure
manifests itself basically through  Pauli Principle, and a very simple expression is
derived  for the neutron and proton induced decays rates,  $\Gamma_n$ and $\Gamma_p$,
which does not involve the  spectroscopic factors.
We use the standard strangeness-changing weak $\Lambda N\to NN$ transition potential which
comprises  the exchange of the complete pseudoscalar and vector meson octets
($\pi,\eta,K,\rho,\omega,K^*$), taking into account some important
parity violating transition operators that  are systematically omitted in the literature.
 The interplay between  different mesons
in the decay of ${^{12}_\Lambda C}$ is  carefully analyzed.
With the commonly used parametrization in  the One-Meson-Exchange Model (OMEM),
the calculated  rate $\Gamma_{NM}= \Gamma_n+\Gamma_p$ is of the order of the free
$\Lambda$ decay rate $ \Gamma^0$($\Gamma_{NM}^{\rm th}\cong \Gamma^0$) and is
consistent with experiments.
Yet, the measurements  of  $\Gamma_{n/p}=\Gamma_n/\Gamma_p$
 and of $\Gamma_p$  are not well accounted for by the
 theory  ($\Gamma_{n/p}^{\rm th}\lsim 0.42; \Gamma_p^{\rm th}\gsim 0.60~\Gamma^0$).
It is suggested that, unless additional degrees of freedom are incorporated,
the OMEM parameters should be radically modified.

\end{abstract}
\pacs{PACS numbers:21.80.+a,21.60.-n,13.75.Ev,25.80.Pw}

\section{Introduction}

 The free $\Lambda$ hyperon weak decay (with transition rate $\Gamma^0= 2.50
\cdot 10^{-6}$ eV) is radically modified  in the nuclear
environment. First, due to Pauli
principle the mesonic decay rate $\Gamma_M\equiv \Gamma
(\Lambda\rightarrow N\pi$) is strongly blocked for $A\ge 4$.
  Second, new nonmesonic (NM) decay channels $\Lambda N\rightarrow NN$
become open, where there are no pions in the final state. The corresponding
transition rates can be stimulated either by protons,
$\Gamma_p\equiv\Gamma(\Lambda p\rightarrow np)$, or by neutrons,
$\Gamma_n\equiv\Gamma(\Lambda n\rightarrow nn)$. The ultimate
result is that in the mass region above
 $A= 12$ the total hypernuclear weak decay rates
$\Gamma_M+\Gamma_{NM}$ ($\Gamma_{NM}=\Gamma_n+\Gamma_p$)
 are almost constant and close to $\Gamma^0$ \cite {Pa00}.

Because of the practical impossibility of having stable $\Lambda$
beams, the NM  decays in hypernuclei offer the best
opportunity to examine the $\Delta S=-1$ nonleptonic weak
interaction between hadrons. Yet, the major motivation for
studying these processes stems from the inability of the present theories to account
for the measurements,
in spite of the huge theoretical effort that has been invested in this issue over several
decades
\cite{Bl63,Ad67,Mc84,Ta85,Co90,Al91,Ra92,Ra95,Pa95,Pa95a,Du96,Pa97,Ra97,Os98,Pa98,In98,It98,Pa99,Alb00,Sa00,Sa00a,Al00,Pa01,Os01,Jun02}.
 More precisely, the theoretical models
reproduce fairly
well the experimental values of the  total width
$\Gamma_{NM}$ ($\Gamma_{NM}^{\rm exp}\cong \Gamma^0$),
 but the ratio $\Gamma_{n/p}\equiv\Gamma_n/\Gamma_p$ ($0.5\le \Gamma_{n/p}^{\rm
exp}\le 2$) remains a puzzle.

In the one meson-exchange  model (OMEM), which is very often used to describe the hypernuclear
 $\Lambda N\rightarrow NN$ decay, it is assumed that the process
 is triggered via the exchange of a virtual meson.
The obvious candidate is the one-pion-exchange (OPE) mechanism,
and following the pioneering investigations of Adams \cite{Ad67}
\footnote{McKellar and Gibson \cite{Mc84} have pointed out that this publication
contains a very  important error and that the decay rates given by
Adams \cite{Ad67} should be multiplied by 6.81.},
several calculations  have been done in $^{12}_\Lambda C$,
yielding: $\Gamma_{NM}^{\rm (OPE)}\cong {\Gamma^0}$ and
$\Gamma_{n/p}^{\rm (OPE)}\cong 0.1-0.2$ \cite{Pa95,Du96,Pa97,Sa00,Pa01}.

The importance of the $\rho$ meson in the weak decay mechanism
was first discussed by McKellar and Gibson \cite{Mc84}. They
 found that, because of
the sensitivity of the results to the unknown $\Lambda N\rho$ vertex,
 the estimates for $\Gamma_{NM}$ could vary by a
factor of 2 or 3 when the potential $V_\rho$ was  included.
(See also Ref. \cite{Ta85}.) The present-day consensus is, however, that
the effect of the $\rho$-meson on both $\Gamma_{NM}$ and $\Gamma_{n/p}$ is small
\cite{Pa95,Du96,Pa97,Os98}.

Until recently  there have been   quite dissimilar opinions
regarding the full OMEM,
which encompasses  all pseudoscalar mesons ($\pi, \eta, K$) and
all vector mesons ($\rho, \omega, K^*$). In fact, while  Dubach \etal \cite{Du96}
claimed that the inclusion of additional exchanges in the
$\pi+\rho$ model  plays a major role in increasing the $n/p$
ratio, Parre\~no, Ramos and Bennhold \cite{Pa97}, and Sasaki, Inoue and Oka \cite{Sa00}
argued that the overall effect of the heavier mesons on this observable was very small.
However,  the two latter  groups have recently corrected
their  calculations  for  a mistake in including
the $K$ and $K^*$ mesons,   and  so their estimates of $\Gamma_{n/p}$  have augmented
quite  substantially \cite{Sa00a,Pa01}. Almost simultaneously, Oset, Jido and  Palomar \cite{Os01} have
also shown that the $K$ meson contribution  was essential to increase $\Gamma_{n/p}$.
However, the experimental data have not been fully explained yet.

In the last few years,  many other attempts  have not been
particularly successful either   in  accounting  for the measured
 $\Gamma_{n/p}$ ratio. To mention just a few of them:
1) analysis of the two-nucleon stimulated process $\Lambda NN\go
NNN$ \cite{Al91,Ra95,Ra97}, 2) inclusion of interaction terms that
violate the isospin ${\Delta}T=1/2$ rule \cite{Pa98,Alb00}, 3)
description of the short range baryon-baryon interaction in terms
of quark degrees of freedom \cite{In98,Sa00}, and 4) introduction
of correlated two-pion exchange potentials besides the OPE
\cite{It98}. Consistent (though not sufficient) increases of the
${n/p}$ ratio were found in the last two works. (For instance,
$\Gamma_{n/p}$  was boosted up to $0.36$ for the decay of
$^{12}_\Lambda C$ \cite{It98}.)
In fact, only  Jun \cite{Jun02} was able  so far to reproduce  well  the $\Gamma_{NM}$,
$\Gamma_{p}$ and $\Gamma_{n/p}$ data.  He has
employed, in addition to the OPE, an entirely phenomenological
4-baryon point interaction for short range interaction, including
the ${\Delta}T=3/2$ contribution as well, and has
conveniently fixed the different model coupling constants.
 Let us
also note that after the present work had been completed, Itonaga
\etal \cite{It02} have updated their studies and  have performed
extensive calculations of the NM decays in the mass region
$4\le A\le 209$, which have revealed that the correlated-$2\pi$
and $1\omega$ exchange potentials significantly improve the
$\Gamma_{n/p}$ ratios over the OPE results.

In the OMEM's, a weak baryon-baryon-meson (BBM)
coupling is always combined with a strong BBM coupling. The  strong one
is determined experimentally with some help from the SU(3)
symmetry, and the involving   uncertainties have been copiously discussed in
the literature \cite{Na77,Ma89,Ha93,Pa94}. It is the weak BBM
couplings which could become  the largest source of errors. In
fact,  only the weak $N\Lambda\pi$ amplitude can be taken from the
experiment, at the expense of  neglecting the off-mass-shell
corrections. All other weak BBM couplings are derived theoretically by
using  $SU(3)$  and $SU(6)_w$ symmetries, octet
dominance, current algebra, PCAC, pole dominance, \etc.
\cite{Co90,Du96,Pa97,De80,Ma69,To82,Ok82,Co83,Na88}.
Assortments of such methods have been developed and employed for a long time
in weak interaction physics to explain the hyperon nonleptonic decays.
Specifically, to obtain the weak BBM couplings for  vector mesons  the $SU(6)_w$
symmetry is used, which is not so well established as the $SU(3)$ symmetry is.
Moreover, the results derived by way of    the $SU(6)_w$ symmetry depend on
the contributions of  factorizable terms $a_V$  and $a_T$, which were
only very roughly estimated \cite{Du96,De80,To82}.
Well aware of all these limitations,  McKellar and Gibson \cite{Mc84} have allowed for
an arbitrary phase between the $\rho$ and the $\pi$ amplitudes in the  $\pi+\rho$ model.
 The same criterion was  adopted by
 Takeuchi, Takaki and Band$\bar{\rm o}$ \cite{Ta85}.

We wish to restate  that the  OMEM transition potential is
purely phenomenological and that it is not derived from a fundamental underlying
form, as happens for instance in the case of electro-magnetic transitions or the
 semileptonic weak decays.
Only the OPE model  is a natural and simple extrapolation of the mesonic decay mechanism of the
$\Lambda$ to the NM process: the weak BBM coupling is identical to that
used in the phenomenological description of the free $\Lambda$, and the strong vertex
is the one traditionally used in describing the $\pi NN$ vertex. The assumption is that
this is a valid approximation, although the pion is off the mass shell.
Accordingly, all modern interpretations of the NM weak decay use the OPE
as the basic building block for the medium and long range part of the
decay interaction. On the contrary, the full OMEM is not used very often and, in place of
the one-meson  $\eta, K,\cdots$ exchanges, other mechanisms are employed as refer to above.
One should also keep in mind that both the strong and weak BBM couplings, as well
the meson masses, can become significantly renormalized by the nuclear environment \cite{Br02}.

The high momentum transfer in the NM decay
makes the corresponding transition amplitude very sensitive to the short
range behavior of the $NN$ and $N\Lambda$ interactions.
In fact, quite recently it has been pointed out that the final state interactions (FSI) have
very large influence on the total and partial decay rates \cite{Pa01}
(see also Ref. \cite{It02}).
\footnote{The FSI  also make  hard the extraction of the $n/p$ ratio
from the experimental data \cite{Mo74,Sz91,No95,Bh98,Ha02}, and to surmount
this difficulty  Hashimoto
 {\sl et al.}\cite{Ha02} have quite recently  combined   the Monte Carlo FSI
 internuclear cascade models from
 Ref. \cite{Ra97} with the geometry of the detectors.  Moreover, Golak \etal \cite{Go99} have
shown  that the FSI, in principle, hinder the measurement of  the ${n/p}$ ratio in
$^{3}_\Lambda H$.}
Due to the same reason one could expect that the nuclear structure effects not included in
the main field (such as the RPA or pairing correlations, higher order seniority excitations
in the initial and final states,\etc) should not play an important role.
Yet,  it could be useful to understand this issue more
genuinely and to get a more complete control on the nuclear structure aspect of the problem.
These are  the main motivations for the present work.

    The only existing shell model framework
    for the hypernuclear decay is the one based on the weak-coupling model (WCM)
between the hyperon and the $(A-1)$ core \cite{Ra92,Pa97,It98,It02}.
It involves the technique of coefficients of fractional parentage,
and the spectroscopic factors (SF)
explicitly   appear in  the expressions for the transition rates.
Yet, in nuclear structure calculations it is in general simpler to evaluate
the transition probabilities directly from the wave functions, instead of
doing it via the SF.
 Here we first develop a fully general shell model formalism and
then we work out thoroughly  the
 extreme particle-hole model (EPHM) and the  Quasiparticle Tamm-Dancoff
 Approximation (QTDA) for the even-mass hypernuclei.

Owing to the above mentioned characteristics of the OMEM it might be legitimate to
ask whether it is possible to account for all three data $\Gamma_{NM}$,
$\Gamma_{p}$ and $\Gamma_{n/p}$ by not fully complying  with
the constraints imposed by
the $SU(3)$, $SU(6)_w$ and chiral symmetries on the BBM couplings.
To find out in which way  should  these  parameters be varied we perform a
 multipole expansion of the transition rate in the framework of the
 EPHM, which unravels in an analytic way the interplay between  different mesons in each
multipole channel.

Attention will be given also to the
parity violating  potential, since there are several typographical errors in
the recent papers \cite{Pa97,Pa99,Pa01}, regarding this part of the
transition potential.  We will also consider  some important  contributions
due to the vector mesons, which, although always  included  in the description of
the nuclear parity violation \cite{De80,Ad85,Ha95,Ha01},
have been so far  neglected in all  studies of the NM hypernuclear decays,
 except those  of Dubach  \etal \cite{Du96,To82}.

The outline of this paper is as follows:
The general shell model formalism for the hypernuclear $\Lambda N\go NN$
 weak decay  is developed in Sec. II.
 The nonrelativistic approximation for
 the effective Hamiltonian is presented in Sec. III. The EPHM and QTDA are
 explained in Sec. IV, where
  the multipole expansion of $\Gamma_{NM}$ is also done. Numerical evaluations
 of the ${^{12}_\Lambda C}\go {^{10}C}+nn$, and ${^{12}_\Lambda C}\go {^{10}B}+pn$
 decay rates, are carried out in Sec. V,
 and the conclusions are presented in Sec. VI.
The  formulae for the nuclear matrix elements are summarized in the Appendix.

\newpage
\section{  Transition Rate}

 The decay rate, of a  hypernucleus (with spin $J_I$ and energy $\E_I$) to
residual nuclei (with  spins $J_F$ and energies $\E_F$) and two free nucleons
(with  total spin $S$ and energies $\epsilon_{p}$ and $\epsilon_{P}$), follows from
Fermi's  Golden  Rule
\begin{equation}
\Gamma=2\pi \sum_{SM_SJ_FM_F} \int |\bra{\pb\Pb SM_S;J_FM_F}
V\ket{J_IM_I}|^2 \delta(\epsilon_{p}+\epsilon_{P}+\E_F -\E_I)
\frac{d{\bf p}}{(2\pi)^3}\frac{d{\bf P}}{(2\pi)^3}.
\label{2.1}\end{equation}
Here, $V$ is  the weak hypernuclear potential,
 the wave functions for the kets $\ket{\pb\Pb SM_S;J_FM_F}$
and $\ket{J_IM_I}$ are assumed to be antisymmetrized  and normalized,
and a transformation to the  relative  and  center of mass (c.m.)
momenta, $\pb$ and $\Pb$,  is already implied, \ie
\be
\pb=\fot(\pb_1-\pb_2),~~~~~\Pb=\pb_1+\pb_2.
\label{2.2}\end{equation}

It is convenient  to define the quantity
\br
\I(p,P)&=&(4\pi)^{-4}\sum_{SM_SJ_FM_F}\int d\Omega_pd\Omega_P
|\bra{\pb\Pb SM_S;J_FM_F}V\ket{J_IM_I}|^2,
\label{2.3}\er
and rewrite \rf{2.1} as:
\be
\Gamma=\frac{16M_N^3}{\pi}
\int_0^{\Delta_F}d\e\sqrt{\e(\Delta_F-\e)}\I(p,P),
\label{2.4}\ee
where $P=2\sqrt{M_N\e}$, $p=\sqrt{M_N(\Delta-\e)}$,  $\Delta_F=\E_I
-\E_F-2M_N$,  and $M_N$ is the nucleon mass.

The partial wave expansion of the wave function  of the
non-antisymmetrized two-particle  ket $|{\Pb\pb SM_S})$ is then performed:
\br
({\rb\Rb s_1s_2}|{\Pb\pb SM_S})
&=&(4\pi)^2\sum_{lmLM}i^{l+L} Y^*_{lm}(\hat{\pb})
Y^*_{LM}(\hat{\Pb}) ({\rb\Rb s_1s_2}|{plm,PLM,SM_S}),
\label{2.5}\er
where
\br
({\rb\Rb s_1s_2}|{plm,PLM,SM_S})
&=&Y_{lm}(\hat{\rb})Y_{LM}(\hat{\Rb})j_{l}(pr)j_{L}(PR)\chi_{SM_S}(s_1s_2),
\label{2.6}\er
describe the spherical free waves for the outgoing particles,
\be
\rb=\rb_1-\rb_2,~~~~~\Rb=\fot(\rb_1+\rb_2).
\label{2.7}\end{equation}
are the relative and  c.m. coordinates, and
$l$ and $L$ are the quantum numbers for the relative (${\bf l}$) and c.m. (${\bf L}$)
orbital angular momenta.
After performing the angular integration in \rf{2.3} we
obtain:
\br
\I(p,P)&=&\sum_{SM_SJ_FM_F}\sum_{lmLM}
|\bra{plm,PLM,SM_S;J_FM_F}V\ket{J_IM_I}|^2,
\label{2.8}\er
which   goes into
\br
\I(p,P)&=&\sum_{SlL\lambda JJ_F}
|\bra{pPlL\lambda SJ,J_F;J_IM_I}V\ket{J_IM_I}|^2.
\label{2.9}\er
when the angular momentum couplings: ${\bf l}+{\bf L}={\mbl}$,
${\mbl}+{\bf S}={\bf J}$ are carried out.
The quantum number $M_I$ is superfluous and will be omitted from now on.

The transition potential is written in the Fock space as: \br
V=\sum_{lL\lambda SJj_Nj_\Lambda} {\bra{pPlL\lambda
SJ}V\ket{j_\Lambda j_NJ}}
\left(a^\dag_{pl\fot}a^\dag_{PL\fot}\right)_{\lambda SJ}\cdot
\left(a_{\overline{j}_N}a_{\overline{j}_\Lambda}\right)_{J},
\label{2.10}\er where, in the same way as in \rf{2.1}, a
transformation to the  relative  and  c.m. momenta is implied.
Here,  $j_\Lambda$ and $j_N$ are the single-particle shell-model
states of the decaying particles, and $a_{\overline{jm}}=(-)
^{j+m}a_{j-m}$ \cite{Boh69}. One gets \br
\I(p,P)&=&\hat{J}_I^{-2}\sum_{SlL\lambda JJ_F^\alpha}
\left|\sum_{j_Nj_\Lambda}{\bra{pPlL\lambda SJ}V\ket{j_\Lambda
j_NJ}} \Bra{J_I}\left( a_{j_N}^\dag
a_{j_\Lambda}^\dag\right)_{J}\Ket{J_F^\alpha}\right|^2,
\label{2.11}\er
 where the transition amplitudes
 $\Bra{J_I}\left( a_{j_N}^\dag a_{j_\Lambda}^\dag\right)_{J}\Ket{J_F}$ are reduced with
 respect to the angular momenta, the label $\alpha$ stands for different final states
 with the same spin $J_F$, and  $\hat{J}\equiv\sqrt{2J+1}$.

The effective weak hypernuclear interaction is isospin dependent, \ie
\be
V(\rb,s_\Lambda s_N,t_\Lambda t_N)=\sum_{\tau=0,1}\V_\tau(\rb,s_\Lambda s_N){\cal T}_\tau,~~~~~
{\cal T}_\tau=\left\{
\begin{array}{c}1\\
  \mbt_\Lambda\cdot\mbt_N
\end{array}\right.,
\label{2.12}\ee and therefore the nuclear matrix elements have to
be evaluated  in the isospin formalism. This  implies that
\rf{2.11} goes into \br \I_{m_{
t_N}}(p,P)&=&\hat{J}_I^{-2}\sum_{S\lambda lLTJJ_F^\alpha}
\nn\\
&\x&\left|\sum_{j_Nj_\Lambda} \M(pPlL\lambda SJT;{j_\Lambda
j_N,m_{t_N}}) \Bra{J_I}\left( a_{j_N m_{ t_N}}^\dag a_{j_\Lambda
}^\dag\right)_{J}\Ket{J_F^\alpha}\right|^2,
\nn\\
\label{2.13}\er
where
\br
\M(pPlL\lambda SJT;{j_\Lambda j_N,m_{t_N}})&=&\sqi [1-(-)^{l+S+T}]
\nn\\
&\x&\sum_{\tau}({pPlL\lambda SJ}|\V_\tau |{j_\Lambda j_N J})
({TM_T}=m_{t_\Lambda}+m_{ t_N}|{\cal T}_\tau| {m_{t_\Lambda}m_{ t_N}}),
 \label{2.14}\er
is the antisymmetrized nuclear matrix element,
and $ m_{t_p}={\small \fot}$ and $m_{t_\Lambda}\equiv m_{t_n}=-{\small \fot}$.
It is assumed, as usual  \cite{Pa97}, that $\Lambda$ behaves as a
$\ket{{\small \fot},-{\small \fot}}$  isospin state.
In that way the  phenomenological $\Delta T={\small \fot}$
rule is incorporated in the effective interaction.
Note that in \rf{2.13} and \rf{2.14} $m_{ t_N}=M_{T}-m_{t_\Lambda}$.

To evaluate $({pPlL\lambda SJ}|\V_\tau |{j_\Lambda j_N J})$
one has to carry out the $jj-LS$ recoupling and  the Moshinsky
transformation \cite{Mo59} on the ket $|{j_\Lambda j_NJ})$ to get
\br
({pPlL\lambda SJ}|\V_\tau |{j_\Lambda j_N J})&=&\hat{j}_\Lambda\hat{j}_N
\sum_{\lambda' S'{\sf nlNL}}
\hat{\lambda'}\hat{S'} \ninj{l_\Lambda}
{\fot}{j_\Lambda}{l_N}{\fot}{j_N}{\lambda'}{S'}{J}
\nn\\
&\x& ({\sf nlNL}\lambda'|n_\Lambda l_\Lambda n_Nl_N\lambda')
(pPlL\lambda SJ|\V_\tau |{\sf nlNL} \lambda' S'J),
\label{2.15}\er
where $(\cdots|\cdots)$ are the Moshinsky brackets \cite{Mo59}.
Here,   ${\sf l}$ and ${\sf L}$ stand for  the quantum numbers of
the  relative and  c.m. orbital angular momenta in the
$\Lambda N$ system.
The explicit expressions for the transition potentials are
given in the next section, and the formulae that are needed to evaluate
the matrix   elements $(pPlL\lambda SJ|\V_\tau |{\sf nlNL} \lambda' S'J)$
and $({TM_T}|{\cal T}_\tau| {m_{t_\Lambda}m_{ t_N}})$ are summarized  in the Appendix.

When the hyperon  is assumed to be weakly coupled to the $A-1$
core, which implies that the interaction of $\Lambda$ with core
nucleons is disregarded, one has that
$\ket{J_I}\equiv\ket{(J_Cj_\Lambda)J_I}$, where   $J_C$ is the
spin of the core. From \br \Bra{J_I}\left( a_{j_N m_{ t_N}}^\dag
a_{j_\Lambda }^\dag\right)_{J}\Ket{J_F}
&=&(-)^{J_F+J+J_I}\hat{J}\hat{J}_I
\sixj{J_C}{J_I}{j_\Lambda}{J}{j_N}{J_F} \Bra{J_C}a_{j_Nm_{
t_N}}^\dag\Ket{J_F}, \label{2.16}\er we obtain \br \I_{m_{
t_N}}(p,P)&=&\sum_{SlL\lambda J TJ_F^\alpha}\hat{J}^{2}
\nn\\
&\x&\left|\sum_{j_N} \M(pPlL\lambda SJT; j_\Lambda j_N,m_{
t_N})\sixj{J_C}{J_I}{j_\Lambda}{J}{j_N}{J_F} \Bra{J_C}a_{j_N m_{
t_N}}^\dag \Ket{J_F^\alpha}\right|^2.
\nn\\
\label{2.17}\er

Occasionally it could be convenient to include the isospin
coupling as well into $\Bra{J_C}a_{j_N m_{ t_N}}^\dag \Ket{J_F}$,
and work with the spin-isospin reduced parentage coefficients \br
\Bra{J_CT_C}|a_{j_N \fot}^\dag |\Ket{J_FT_F}&=&\hat{T}_C
\frac{\Bra{J_CT_CM_{T_C}}a_{j_N m_{ t_N}}^\dag
\Ket{J_FT_FM_{T_F}}} {(T_FM_{T_F}\fot m_{ t_N}|T_CM_{T_C})},
\label{2.18}\er where $T_C,M_{T_C}$ and $T_F,M_{T_F}$ are the
isospin quantum numbers of the core and residual nuclei,
respectively. In this case \br \I_{m_{
t_N}}(p,P)&=&\hat{T}_C^{-2}\sum_{ J_F^\alpha T_FSlL\lambda
JT}\hat{J}^{2} (T_FM_{T_F}\fot m_{ t_N}|T_CM_{T_C})^2
\nn\\
&\x&\left|\sum_{j_N} \M(pPlL\lambda SJ;{j_\Lambda j_N,m_{
t_N}})\sixj{J_C}{J_I}{j_\Lambda}{J}{j_N}{J_F} \Bra{J_CT_C}|a_{j_N
\fot}^\dag |\Ket{J_F^\alpha T_F}\right|^2.
\nn\\
\label{2.19}\er

Thus, knowing the transition potential $V$ and the initial and final nuclear
wave functions $\ket{J_I}$ and $\ket{J_F}$ (or $\ket{J_C}$ and $\ket{J_F}$),
we can evaluate the transition rate \rf{2.4}, with the integrations  going up to
\be
\Delta_{j_Nm_{ t_N}}=M_\Lambda-M_N+\e_{j_\Lambda}+\e_{j_Nm_{ t_N}},
\label{2.20}\ee
where  $\e_{j_\Lambda}$ and $\e_{j_Nm_{ t_N}}$ are the single particle energies.

\section{Effective Interaction}


As the reduction of the relativistic one-meson exchange t-matrix, to the
nonrelativistic effective potential $V$, is in the literature
\cite{Mc84,Co90,Pa95,Pa95a,Du96,Pa97,Pa01,To82,Na88} it
will not be repeated here. For the parity conserving (PC) potential we will
just list a few results that are indispensable for establishing the notation and
for the final  discussion.  More attention will be given  to the
parity violating (PV) potentials.
In  dealing  with them  some tricky details  appear concerning
the passage from the momentum space to the coordinate  space.
We first illustrate the procedure for one pseudoscalar meson ($\pi$)  and
one  vector meson ($\rho$), and afterwards  we generalize the results  to all six mesons.


The effective strong (S) and    weak (W) Hamiltonians read \br
\H^S_{N N\pi
}&=&ig_{NN\pi}\bar{\psi}_N\gamma_5\mbpi\cdot\mbt\psi_N,
\nn\\\nn\\
\H^S_{N N\rho }&=&\bar{\psi}_N
\left(g^V_{NN\rho}\gamma^\mu+ig^T_{NN\rho}\frac{\sigma^{\nu\mu}\partial_\nu}{2M}\right)
\mbr_\mu\cdot\mbt\psi_N,
\nn\\\nn\\
\H^W_{\Lambda N\pi
}&=&iG_Fm_\pi^2\bar{\psi}_N\left(A_\pi+B_\pi\gamma_5\right)
\mbpi\cdot\mbt\psi_\Lambda\bin{0}{1},
\nn\\\nn\\
\H^W_{\Lambda N\rho}&=&G_Fm_\pi^2\bar{\psi}_N
\left(A_\rho\gamma^\mu\gamma_5+B^V_\rho\gamma^\mu+iB_\rho^T\frac{
\sigma^{\nu\mu}\partial_\nu} {2\overline{M}}\right)
\mbr_\mu\cdot\mbt\psi_\Lambda\bin{0}{1}, \label{3.1}\er where
$G_Fm_\pi^2$ is the weak coupling constant,   $\psi_N$ and
$\psi_\Lambda$ are the baryon fields, $\mbpi$ and $\mbr$ are the
meson fields, $\mbt$ is the isospin operator, $M$ the nucleon
mass, and $\overline{M}$ the average between the nucleon and
$\Lambda$ masses. The isospin spurion $\bin{0}{1}$ is included in
order to enforce the empirical $\Delta T={\small \fot}$ rule
\cite{Pa97}.


 The corresponding nonrelativistic t-matrix
in the momentum space (with the hyperon $\Lambda$  being always
in the first vertex) is:
\newpage
\br t_\pi(\qb)&=&-\mbt_\Lambda\cdot\mbt_N
\frac{\A_\pi(\mbs_N\cdot\qb) +
\B_\pi(\mbs_\Lambda\cdot\qb)(\mbs_N\cdot\qb)} {m_\pi^2+\qb^2},
\nn\\\nn\\
t_\rho(\qb,\Qb)&=&-\mbt_\Lambda\cdot\mbt_N
\frac{i\A_\rho(\mbs_\Lambda\x\mbs_N)
\cdot\qb-2\A'_\rho\mbs_\Lambda\cdot\Qb
+\B_\rho(\mbs_\Lambda\x\qb)(\mbs_N\x\qb)-\B'_\rho}
{m_\rho^2+\qb^2}, \label{3.2}\er where the coupling constants
$\A_M$, $\A'_M$, $\B_M$ and  $\B'_M$ are defined in Table
\ref{table1}, and \br \qb&=&\pb'-\pb;~~~~\Qb=\fot(\pb'+\pb),
\label{3.3}\er with $\pb'$ and $\pb$ being, respectively,  the
relative momenta for the initial and final states. (We have
adopted this labeling to be consistent with \rf{2.2}.) In the
momentum space the  potential reads: \br
\bra{\pb_1\pb_2}V\ket{\pb'_1\pb'_2}&=&
-(2\pi)^3\delta(\pb'_1+\pb'_2-\pb_1-\pb_2)t(\qb,\Qb),
\label{3.4}\er and in order to arrive to the coordinate space the
Fourier transform is applied: \br
\bra{\rb_1\rb_2}V\ket{\rb'_1\rb'_2}&=&
\int\frac{d\pb'_1}{(2\pi)^3}\frac{d\pb'_2}{(2\pi)^3}\frac{d\pb_1}{(2\pi)^3}
\frac{d\pb_2}{(2\pi)^3}\bra{\pb_1\pb_2}V\ket{\pb'_1\pb'_2}
\nn\\\nn\\
&\x&\exp\{i[\pb'_1\cdot\rb'_1+\pb'_2\cdot\rb'_2-\pb_1\cdot\rb_1-\pb_2\cdot\rb_2]\}.
\label{3.5}\er After some trivial integrations and  the coordinate
transformation: \br \xb&=&\rb-\rb';~~~~\Xb=\fot(\rb'+\rb),
\label{3.6}\er we get \br \bra{\rb_1\rb_2}V\ket{\rb'_1\rb'_2}&=&
-\delta(\Rb'-\Rb)\int \frac{d\Qb}{(2\pi)^3} \frac{d\qb}{(2\pi)^3}
\exp[i(\Qb\cdot\xb+\qb\cdot\Xb)]t(\qb,\Qb). \label{3.7}\er To
carry out the integration on  $\qb$ and $\Qb$ we make use of the
result:
\newpage
\br &&\int \frac{d\Qb}{(2\pi)^3}\frac{d\qb}{(2\pi)^3} \qb\frac{
e^{i(\Qb\cdot\xb+\qb\cdot\Xb)}}{m_M^2+\qb^2}
=-i\delta(\rb'-\rb)\fb_M^{(-)}(r),
\nn\\\nn\\
&&\int \frac{d\Qb}{(2\pi)^3}\frac{d\qb}{(2\pi)^3} \Qb\frac{
e^{i(\Qb\cdot\xb+\qb\cdot\Xb)}}{m_M^2+\qb^2}
=\frac{i}{2}\delta(\rb'-\rb)\fb_M^{(+)}(r), \nn\\\label{3.8}
&&\int \frac{d\Qb}{(2\pi)^3}\frac{d\qb}{(2\pi)^3}
(\mbs_1\cdot\qb)(\mbs_2\cdot\qb)\frac{
e^{i(\Qb\cdot\xb+\qb\cdot\Xb)}}{m_M^2+\qb^2}
=-\delta(\rb'-\rb)[f^S_M(r)(\mbs_1\cdot\mbs_2)+f^T_M(r)S_{12}(\rh)],
\nn\\\nn\\
&&\int \frac{d\Qb}{(2\pi)^3}\frac{d\qb}{(2\pi)^3}
(\mbs_1\x\qb)(\mbs_2\x\qb)\frac{
e^{i(\Qb\cdot\xb+\qb\cdot\Xb)}}{m_M^2+\qb^2}
=-\delta(\rb'-\rb)[2f^S_M(r)(\mbs_1\cdot\mbs_2)-f^T_M(r)S_{12}(\rh)],
\nn\\\er where \br
S_{12}(\rh)&=&3(\mbs_1\cdot\hat{r})(\mbs_2\cdot\hat{r})
-(\mbs_1\cdot\mbs_2) =\sqrt{\frac{24\pi}{5}}Y_2(\rh)
\cdot[\mbs_1\x\mbs_2]_2, \label{3.9}\er is the tensor operator,
and the radial dependence is contained in: \br
\fb_M^{(-)}(r)&=&\left[\mbn,f_M(r)\right]=\mbn f_M(r) \equiv
\rh\frac{\partial}{\partial r}f_M(r)= \rh  f'_M(r),
\nn\\\nn\\
\fb_M^{(+)}(r)&=&\left\{\mbn,f_M(r)\right\}=\mbn f_M(r)+2f_M(r)\mbn,
\nn\\\nn\\
f_M^{S}(r)&=&\frac{1}{3}\left[m_M^2f_M(r) -\delta(\rb)\right],
\nn\\\nn\\
f_M^T(r)&=&\frac{m_M^2}{3}
\left[1+\frac{3}{m_M r}+\frac{3}{(m_M r)^2}\right]f_M(r),
\label{3.10}\er
with $\mbn\equiv\mbn_{12}=\mbn_1=-\mbn_2$, and
\br
f_M(r)&=&\frac{e^{-m_M r}}{4\pi r};~~~r=|\rb_1-\rb_2|,
\nn\\\nn\\
 f'_M(r)&=&-m_M \left(1+\frac{1}{m_Mr}\right)f_M(r).
\label{3.11}\er
Thus \rf{3.7} reads
\br \bra{\rb_1\rb_2}V\ket{\rb'_1\rb'_2}&=&
\delta(\rb'-\rb)\delta(\Rb'-\Rb)V(\rb), \label{3.12}\er where the
transition potential for the $\pi+\rho$ model is: \br
V_{\pi+\rho}(\rb)&=&\mbt_\Lambda\cdot\mbt_N \left\{
(\mbs_\Lambda\cdot\mbs_N)\left[\B_\pi f_\pi^S(r)+2\B_\rho
f_\rho^S(r)\right]
+S_{\Lambda N}(\rh)\left[\B_\pi f_\pi^T(r)-\B_\rho f_\rho^T(r)\right]
\right.\nn\\\nn\\
&+&\left.\B'_\rho f_\rho(r)-i\A_\pi\mbs_N\cdot  \fb_\pi^{(-)}(r)-i\A'_\rho\mbs_\Lambda\cdot\fb_\rho^{(+)}(r)
+\A_\rho (\mbs_\Lambda\x\mbs_N)\cdot\fb_\rho^{(-)}(r)\right\}.
\nn\\
\label{3.13}\er

The complete  potential  can now be cast in the form
\rf{2.12}, with the isoscalar ($\eta,\omega$) and isovector ($\pi,\rho$) mesons
  giving rise to $\V_0$ and $\V_1$, respectively,
while the strange mesons ($K,K^*$) contribute to both.
We get:
\br
\V^{PV}_0(\rb,s_1s_2)&=&-i \mbs_N\cdot\left[ \A_\eta\fb_\eta^{(-)}(r)
-\A'_{K^*_0}\fb_{K^*}^{(+)}(r)\right]
+i\mbs_\Lambda\cdot\left[ \A_{K_0}\fb_K^{(-)}(r)-\A'_\omega\fb_\omega^{(+)}(r)\right]
\nn\\\nn\\
&+&(\mbs_\Lambda\x\mbs_N)\cdot\left[\A_\omega \fb_\omega^{(-)}(r)+
\A_{K^*_0}\fb_{K^*}^{(-)}(r)\right],
\nn\\\nn\\
\V^{PV}_1(\rb,s_1s_2)&=&-i\mbs_N\cdot [\A_\pi \fb_\pi^{(-)}(r)-\A'_{K^*_1}\fb_{K^*}^{(+)}(r)]
+i\mbs_\Lambda\cdot\left[ \A_{K_1}\fb_K^{(-)}(r)-\A'_\rho\fb_\rho^{(+)}(r)\right]
\nn\\\nn\\
&+&(\mbs_\Lambda\x\mbs_N)\cdot\left[\A_\rho \fb_\rho^{(-)}(r)+
\A_{K^*_1}\fb_{K^*}^{(-)}(r)\right],
\label{3.14}\er
for the PV potential, and
\newpage
\br
\V^{PC}_0(\rb,s_1s_2)&=&(\mbs_\Lambda\cdot\mbs_N)\left[\B_\eta f_\eta^S(r)+\B_{K_0} f_K^S(r)
+2\B_\omega f_\omega^S(r)+2\B_{K^*_0} f_{K^*}^S(r)\right]
\nn\\\nn\\
&+&S_{\Lambda N}(\rh)\left[\B_\eta f_\eta^T(r)+\B_{K_0} f_{K}^T(r)
-\B_\omega f_\omega^T(r)-\B_{K^*_0} f_{K^*}^T(r)\right]
\nn\\\nn\\
&+&\B'_\omega f_\omega(r)+\B'_{K^*_0} f_{K^*}(r),
\nn\\\nn\\
\V^{PC}_1(\rb,s_1s_2)&=&(\mbs_\Lambda\cdot\mbs_N)\left[\B_\pi f_\pi^S(r)+\B_{K_1} f_K^S(r)
+2\B_\rho f_\rho^S(r)+2\B_{K^*_1} f_{K^*}^S(r)\right]
\nn\\\nn\\
&+&S_{\Lambda N}(\rh)\left[\B_\pi f_\pi^T(r)+\B_{K_1} f_{K}^T(r)
-\B_\rho f_\rho^T(r)-\B_{K^*_1} f_{K^*}^T(r)\right]
\nn\\\nn\\
&+&\B'_\rho f_\rho(r)+\B'_{K^*_1} f_{K^*}(r),
\label{3.15}\er
for the PC potential.
The overall coupling constants $\A_M$, $\A'_M$, $\B_M$ and , $\B'_M$,
are listed in Table \ref{table1}, with the weak couplings  for kaons defined as:
\begin{table}[h]
\caption{Isoscalar ($\tau=0$) and isovector ($\tau=1$) coupling constants   in units of
${G_Fm_\pi^2}=2.21\x 10^{-7}$. \label{table1}}
\begin{tabular}{|c|ccccc|}
M&&$\A_M$&$\A'_M$&$\B_M$&$\B'_M$\\
\hline
\hline
&&    &    $\tau=0$ &       &         \\
\hline
$\eta$
&&$A_\eta \displaystyle\frac{g_{NN\eta}}{2M}$
&$   - $
&$\displaystyle\frac{B_\eta}{2M}\displaystyle \frac{ g_{NN\eta}}{2\overline{M}}$
&$  -$
\\
&   & &     &       &         \\
$K$
 &           &$A_{K_0}\displaystyle\frac{g_{\Lambda NK}}{2M}$
&$   - $
&$\displaystyle\frac{B_{K_0}}{2M}\displaystyle\frac{g_{\Lambda NK}}{2\overline{M}}$
            &$  -$
            \\
&    &&     &       &         \\
$\omega$
&&$A_\omega\displaystyle\frac{ g^{\sss V}_{NN\omega}+g^{\sss T}_{NN\omega}}{2M}$
&$2A_\omega\displaystyle\frac{ g^{\sss V}_{NN\omega}}{2M}$
&$\displaystyle\frac{B^{\sss V}_\omega+B^{\sss T}_\omega}{2\overline{M}}
\displaystyle\frac{ g^{\sss V}_{NN\omega}
+ g^{\sss T}_{NN\omega}}{2M}$
& ${B^{\sss V}_\omega g^{\sss V}_{NN\omega}}$
\\
& &   &     &       &         \\
$K^*$
&&$A_{K^*_0}\displaystyle\frac{g^{\sss V}_{\Lambda NK^*}+g^{\sss T}_{\Lambda NK^*}}{2M}$
&$2A_{K^*_0}\displaystyle\frac{g^{\sss V}_{\Lambda NK^*}}{2M}$
&$\displaystyle\frac{B^{\sss V}_{K^*_0}+B^{\sss T}_{K^*_0}}{2M}\frac{g^{\sss V}_{\Lambda NK^*}
+g^{\sss T}_{\Lambda NK^*}}{2\overline{M}}$
&$B^{\sss V}_{K^*_0}g^{\sss V}_{\Lambda NK^*} $
\\
& &   &     &       &         \\
\hline
&&    &    $\tau=1$ &       &         \\
\hline
$\pi$
&&$A_\pi\displaystyle\frac{ g_{NN\pi}}{2M}$
&$   - $
&$\displaystyle\frac{ B_\pi}{2M}\displaystyle\frac{ g_{NN\pi}}{2\overline{M}}$
&$  -$
            \\
&  &  &     &       &         \\
$K$
  &         &$2A_{K_1}\displaystyle\frac{g_{\Lambda NK}}{2M}$
&$   - $
&$\displaystyle\frac{ B_{K_1}}{2M}\displaystyle\frac{g_{\Lambda NK}}{2\overline{M}}$
           &$  -$
           \\
&    &&     &       &         \\
$\rho$
&&$A_\rho\displaystyle\frac{ g^{\sss V}_{NN\rho}+g^{\sss T}_{NN\rho}}{2M}$
&$2A_\rho\displaystyle\frac{ g^{\sss V}_{NN\rho}}{2M}$
&$\displaystyle\frac{B^{\sss V}_\rho+B^{\sss T}_\rho}{2\overline{M}}
\displaystyle\frac{ g^{\sss V}_{NN\rho}
+ g^{\sss T}_{NN\rho}}{2M}$
& ${B^{\sss V}_\rho g^{\sss V}_{NN\rho}}$
\\
&&    &     &       &         \\
&&    &     &       &         \\
$K^*$
&&$A_{K^*_1}\displaystyle\frac{g^{\sss V}_{\Lambda NK^*}+g^{\sss T}_{\Lambda NK^*}}{2M}$
&$2A_{K^*_1}\displaystyle\frac{g^{\sss V}_{\Lambda NK^*}}{2M}$
&$\displaystyle\frac{B^{\sss V}_{K^*_1}+B^{\sss T}_{K^*_1}}{2M}\frac{g^{\sss V}_{\Lambda NK^*}
+g^{\sss T}_{\Lambda NK^*}}{2\overline{M}}$
&$B^{\sss V}_{K^*_1}g^{\sss V}_{\Lambda NK^*} $
\\
&&    &     &       &         \\
\end{tabular}
\end{table}

\br
A_{K_0}&=&\frac{C_K^{PV}}{2}+D_K^{PV};~~~~A_{K_1}=\frac{C_K^{PV}}{2},
\nn\\\nn\\
B_{K_0}&=&\frac{C_K^{PC}}{2}+D_K^{PC};~~~~B_{K_1}=\frac{C_K^{PC}}{2},
\nn\\\nn\\
A_{K^*_0}&=&\frac{C_{K^*}^{PV}}{2}+D_{K^*}^{PV};~~~~A_{K^*_1}=\frac{C_{K^*}^{PV}}{2},
\nn\\\nn\\
B^V_{K^*_0}&=&\frac{C_{K^*}^{PC,V}}{2}+D_{K^*}^{PC,V};~~~~B^V_{K^*_1}=\frac{C_{K^*}^{PC,V}}{2},
\nn\\\nn\\
B^T_{K^*_0}&=&\frac{C_{K^*}^{PC,T}}{2}+D_{K^*}^{PC,T};~~~~B^T_{K^*_1}=\frac{C_{K^*}^{PC,T}}{2}.
\label{3.16}\er
The $C$'s and $D$'s are given in Ref. \cite{Pa97}. The operators that have been habitually
omitted in $\V^{PV}(\rb,s_1s_2)$ are those that are proportional to $\A'_M$.

\newpage
\section{Nuclear Models and Multipole Expansion}
\subsection{Extreme Particle-Hole Model}
\begin{figure}[h]
\begin{center}
\vspace{3.1cm}
    \leavevmode
   \epsfxsize = 9cm
     \epsfysize = 8cm
  \epsffile{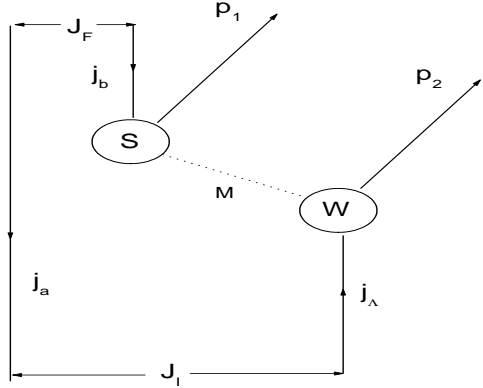}
   \end{center}
\vspace{-3.cm}
 {\tighten \caption{Diagramatic representation of the
 hypernuclear NM weak decay, from the
1p1h state $\ket{j_\Lambda j^{-1}_a;J_I}$ to the
2h state $\ket{j_a^{-1}j_b^{-1}J_F}$, while two nucleons with momenta
$\pb_1$ and $\pb_2$ are emitted into the continuum. S and W are the strong
and the weak vertices, respectively, and M is a non-strange meson.
}} \label{fig1}
\end{figure}

The simplest nuclear shell model is the EPHM,  in which: the
hypernucleus  $^{A}_\Lambda Z$ is described as a $\Lambda$-hyperon
in the single particle state $\ket{j_\Lambda}$ and a hole state
$\ket{{j_a}^{-1}}$ relative to the $^{A}Z$ core, while the
residual  $^{A-2}Z$  and  $^{A-2}(Z-1)$ nuclei are represented by
the two hole states $\ket{{j_a}^{-1}{j_b}^{-1}}$ with respect to
the same core. As illustrated in Fig. 1 $\ket{J_I}\go
\ket{j_\Lambda j^{-1}_a;J_I}$, $\ket{J_F}\go
\ket{{j_a}^{-1}{j_b}^{-1};J_F}$ and $\ket{J_C}\go
\ket{{j_a}^{-1}}$. The parentage coefficients in \rf{2.17} read
\br &&\Bra{J_C}a_{j_bm_{t_{b}}}^\dag\Ket{J_F}= (-)
^{J_F+j_a+j_b}\sqrt{1+\delta_{ab}}\hat{J}_F. \label{N.1}\er In
particular, for $^{12}_\Lambda C$ the initial state is \be
\left(a_{j_\Lambda}^\dag \overline{a}_{j_a} \right)_{J_I}\ket{0}
\equiv\ket{j_\Lambda
j^{-1}_a;J_I}=\ket{1s_{1/2}\Lambda,1p_{3/2}n^{-1};1},
 \label{N.2}\ee
and the final states are:
\br
&&
\left( \overline{a}_{j_a}\overline{a}_{j_b} \right)_{J_F}\ket{0}
\equiv\ket{{j_a}^{-1}{j_b}^{-1};J_F}=\left\{
\begin{array}{ll}
 \ket{(1p_{3/2}n^{-1})^2;0,2},
\ket{1p_{3/2}n^{-1}1s_{1/2}n^{-1};1,2},
\\\\
\ket{(1p_{3/2}n^{-1}1p_{3/2}p^{-1});0,1,2,3},
\ket{1p_{3/2}n^{-1}1s_{1/2}p^{-1};1,2},
\\
\end{array}\right.
\label{N.3}\er
for $\Lambda n\rightarrow nn$ and $\Lambda p\rightarrow np$, respectively. Here $\ket{0}$
is the $^{12}C$ particle vacuum.
As there is only one hole state for each parity, the parentage coefficients with
different $j_b=j_N$ do not interfere among themselves. After summing up on the final states
the integrand \rf{2.17} can be cast in the form
\br
&&\I_{m_{t_b}}(p,P)=\sum_{j_bJ}F_{m_{t_b}J}^{j_b}(p_{3/2})\sum_{S\lambda lLT}
\M^2(pPlL\lambda SJT;{j_\Lambda j_b,m_{t_b}}),
\label{N.4}\er
where
\br
F_{m_{t_b}J}^{j_b}(j_a)&=&\hat{J}^2\hat{j}_b^{-2}\sum_{ J_F=|j_a-j_b|}^{j_a+j_b}
\left[1+(-)^{J_F}\delta_{j_aj_b}\delta_{m_{t_b}m_{t_a}}\right]\hat{J}_F^2
\sixj{j_a}{j_b}{J_F}{J}{J_I}{j_\Lambda}^2.
\label{N.4a}\er
are geometrical factors  which come from the Pauli principle.
Their explicit values for the $1s_{1/2},1p_{3/2},1p_{1/2}$ are listed in Table \ref{table2}.

\begin{table}[h]
\caption{Geometrical   factors $\hat{j}_b^{2}F^{j_b}_{m_{t_bJ}}(j_a)$
\label{table2}}
\begin{center}
\begin{tabular}{|ccccc|}
 $j_a$ &$j_b$& J & $neutrons$& $protons$ \\
\hline
 $1p_{3/2}$& $1s_{1/2}$  & $0$ &$1$ &$1$ \\
&  & $1$ &$3$ &$3$ \\
$1p_{1/2}$& $1s_{1/2}$& $0$ &$1$ &$1$ \\
 && $1$ &$3$ &$3$ \\
\hline
 $1p_{3/2}$& $1p_{3/2}$  & $1$ &$7$ &$6$ \\
& & $2$ &$5$ &$10$ \\
 $1p_{1/2}$& $1p_{3/2}$& $1$ &$6$ &$6$ \\
&& $2$ &$10$ &$10$ \\
\hline
  $1p_{3/2}$& $1p_{1/2}$  & $0$ &$1$  &$1$\\
 & & $1$ &$3$ &$3$ \\
$1p_{1/2}$& $1p_{1/2}$& $0$ &$0$ &$1$\\
 && $1$ &$2$ &$3$ \\
 \end{tabular}
\end{center}
\end{table}

\subsection{Beyond  Extreme Particle-Hole Model}

The EPHM can be straightforwardly improved by going to the quasiparticle
representation. In fact, for all even-mass hypernuclei the initial and final states
can be expressed as:
\br
\ket{J_I}&=&\sum_{j_\Lambda j_a}C_{j_\Lambda j_a}
\left(a_{j_\Lambda}^\dag b^\dag _{j_a} \right)_{J_I}\ket{BCS},
\nn\\
\ket{J_F^\alpha}&=&\sum_{j_a j_b}C_{j_a j_bJ_F}^\alpha
\left(b_{j_a}^\dag b^\dag _{j_b}\right)_{J_F}\ket{BCS},
 \label{M.1}\er
where $b_{j}^\dag=u_j a_{j}^\dag-v_ja_{\overline{j}}$ is the quasiparticle creation
 operator \cite{Ri80}, $\ket{BCS}$ is the BCS vacuum, $j_a$ is always a neutron
 state, while $j_b$ can be both  a neutron and a proton orbital.
 Note that because of the lack of hyperon-hole states, the backward going RPA
 contributions do not appear and one has to work within the QTDA. From \rf{2.13} we get
\br &&\I_{m_{ t_b}}(p,P)=\sum_{lL\lambda SJJ_F^\alpha
T}\hat{J}^{2}\hat{J}_F^{2}
\nn\\
&\x&\left|\sum_{j_\Lambda j_aj_b } (-)
^{j_a+j_b}\sqrt{1+\delta_{ab}}C^\alpha_{j_aj_bJ_F}C_{j_aj_\Lambda}v_{j_b}
\M(pPlL\lambda SJT; j_\Lambda j_b,m_{
t_b})\sixj{j_a}{J_I}{j_\Lambda}{J}{j_b}{J_F} \right|^2.
\nn\\
\label{M.2}\er The residual interaction in the final nuclei
redistributes the transition rates among the states with the same
spin and parity. But, as the NM decay is an inclusive process, \ie
the partial transition rates are summed up coherently over all
final states, such a rearrangement  plays only a very minor role
on the total rates. (The same happens, for instance, in the
neutrino-nucleus reactions and in the $\mu$ meson capture
\cite{Krm02}.) Therefore, it is justifiable to approximate the
final  wave functions by their unperturbed forms, \ie
$\ket{J_F^\alpha}\equiv \left(b_{j_a}^\dag b^\dag
_{j_b}\right)_{J_F}\ket{BCS}$
 and $C^\alpha_{j_aj_bJ_F}\equiv\delta_{\alpha,{j_aj_b}}$.
If, in addition, one assumes that the hyperon is always in the lowest $1s_{1/2}$ state
the last equation takes the form of \rf{N.4}, \ie
\br
&&\I_{m_{t_b}}(p,P)=\sum_{j_aj_bJ}F_{m_{t_b}J}^{j_b}(j_a)
C^{2}_{j_aj_\Lambda}v^{2}_{j_b}\sum_{S\lambda lLT}
\M^2(pPlL\lambda SJT;{j_\Lambda j_b,m_{t_b}}),
\label{M.3}\er
Only
the orbitals $1s_{1/2}$, $1p_{3/2}$ and $1p_{1/2}$ will be used.
In this case, as  seen from Table \ref{table2},
 $F_{pJ}^{j_b}(p_{1/2})=F_{pJ}^{j_b}(p_{3/2})$,
which implies that in the case of protons  the summation on $j_a$   can be performed
analytically. Thus, as
$\sum_{j_a}C^{2}_{j_aj_\Lambda}=1$, one finds out that $\Gamma_p$ does not depend
 at all on the initial wave function. From the same table one also finds out that
$F_{nJ}^{j_b}(j_a)=F_{pJ}^{j_b}(j_a)$, except when
$j_b=j_a$. So, one can expect as well only a weak  dependence
of $\Gamma_n$ on $\ket{J_I}$. This fact is verified  numerically later on.

In summary, we end up with a very simple result for the
transition rates:
\br
\Gamma_{m_{t_b}}&=& \sum_{j_bJ}v^{2}_{j_b}F_{m_{t_b}J}^{j_b}(j_a)
\R_{m_{t_b}J}^{j_b},
\label{M.4}\er
where
\br
\R_{m_{t_b}J}^{j_b}=\frac{16M_N^3}{\pi}\int_0^{\Delta_{j_b}}d\e\sqrt{\e(\Delta_{j_b}-\e)}
\sum_{SlL\lambda T}
\M^2(pPlL\lambda SJT;{j_\Lambda j_b,m_{t_b}}).
\label{M.5}\er
Clearly, the EPHM is contained in \rf{M.4} with the occupation numbers  $v_{j_b}$
equal to one  for the occupied states and to zero for the empty states.

\subsection{Multipole Expansion}

    The EPHM is particularly suitable for performing  the multipole expansion of
    the integrands $\I_{m_{t_b}}$. Thus
we  carry out both the Racah  algebra  in \rf{2.15} and the summations
indicated in \rf{N.4}, keeping in mind that the allowed
quantum numbers $\{{\sf lL}\}$  are:
 $\{00\}$  for the $s_{1/2}$ state, and  $\{01\}$ and $\{10\}$ for the $p_{3/2}$ state.
To simplify the  results we take advantage of the relations
\br
(P0|10)&=&\left(\frac{\pi}{2}\right)^{1/4}b^{3/2}e^{-(Pb)^2/4},
\nn\\
(P1|11)&=&\frac{1}{\sqrt{3}}\left(\frac{\pi}{2}\right)^{1/4}b^{5/2}Pe^{-(Pb)^2/4},
\label{N.5}\er
for the radial integrals $(PL|{\sf{NL}})$ defined in
\rf{A2},  and introduce the ratio
\be
R=\frac{(bP)^2}{3}\equiv\left[\frac{(P1|11)}{(P0|10)}\right]^2,
\label{N.6}\ee
which allows us to work only  with the ${\sf L}=0$ overlap $(P0|10)$. Thus, from now on the label
${\sf L}$ will be disregarded, and to identify  the $s_{1/2}$
and $p_{3/2}$ pieces of the ${\sf l}=0$ strength
we will  use  the ratio $R$, which appears only in the last term of \rf{N.4}.
The results of the multipole expansion for both PC and PV potentials are displayed below.

\subsubsection{Parity conserving contributions}

The matrix elements of the PC  operators $f_M(r)$, $f^S_M(r)(\mbs_\Lambda\cdot\mbs_N)$,
  and $f^T_M(r)S_{\Lambda N}(\rh)$, given by \rf{A1}, can be expressed by means
  of the radial matrix elements \rf{A2} and \rf{A3}, or more precisely through
the moments
\br
{\sf C}_M^{{\sf l}}(p,P)&=&\B'_M(p{\sf l}|f_M|1{\sf l})(P0|10),
\nn\\
{\sf S}_M^{{\sf l}}(p,P)&=&\B_M{(p{\sf l}|f^S_M|1{\sf l})}(P0|10)\x\left\{
\begin{array}{rcc}
  1&\mbox{for}&\pi,\eta,K\\
  2&\mbox{for}&\rho,\omega,K^*
\end{array}
\right.,\nn\\
{\sf T}_M^{l{\sf l}}(p,P)&=&\B_M{(pl|f^T_M|1{\sf l})}(P0|10)\x\left\{
\begin{array}{rcc}
  1&\mbox{for}&\pi,\eta,K\\
  -1&\mbox{for}&\rho,\omega,K^*
\end{array}
\right..
\label{N.8}\er
Introducing the notation:
\br
\begin{array}{ccc}
 \underline{ \tau=0 }&& \underline{\tau=1} \\
 {\sf C}_0= {\sf C}_\omega+{\sf C}_{K_0}
  &;& {\sf C}_1= {\sf C}_\rho+{\sf C}_{K^*_1}, \\
{\sf S}_0= {\sf S}_\eta+{\sf S}_\omega +{\sf S}_{K_0}  +{\sf
S}_{K^*_0}
  &;& {\sf S}_1= {\sf S}_\pi+{\sf S}_\rho+{\sf S}_{K_1}  +{\sf S}_{K^*_1}, \\
{\sf T}_0= {\sf T}_\eta+{\sf T}_\omega+{\sf T}_{K_0}  +{\sf
T}_{K^*_0}
  &;& {\sf T}_1= {\sf T}_\pi+{\sf T}_\rho+{\sf T}_{K_1}  +{\sf T}_{K^*_1}, \\
\end{array}
\label{N.9}\er
for the isoscalar ($\tau=0$) and the  isovector ($\tau=1$) matrix elements,  one gets:
\br
{\cal I}_p&=&2\left(1+R\right)
\left[3\left({\sf S}^0_0\right)^2+9\left({\sf S}^0_1\right)^2
+       \left({\sf C}^0_0\right)^2+7\left({\sf C}^0_1\right)^2
+6\left(3{\sf T}^{20}_1-{\sf T}^{20}_0\right)^2\right.\nn\\
&-&\left.
4{\sf C}^0_0{\sf C}^0_1+12{\sf C}^0_1{\sf S}^0_1
-6{\sf C}^0_0{\sf S}^0_1-6{\sf C}^0_1{\sf S}^0_0
\right]
\nn\\
&+&6\left({\sf S}^1_0\right)^2+42\left({\sf S}^1_1\right)^2
-24{\sf S}^1_0{\sf S}^1_1
+2\left({\sf C}^1_0\right)^2+6\left({\sf C}^1_1\right)^2
-24{\sf C}^1_1{\sf S}^1_1
\nn\\
&+&12{\sf C}^1_1{\sf S}^1_0+12{\sf C}^1_0{\sf S}^1_1
+\frac{6}{5}\left({\sf T}^{11}_0+{\sf T}^{11}_1\right)^2
+\frac{54}{5}\left({\sf T}^{31}_0+{\sf T}^{31}_1\right)^2,
\label{N.10}\er
for the decay $\Lambda p\rightarrow np$, and
\br
{\cal I}_n&=&\left(1+\frac{7R}{3}\right)
\left(3{\sf S}^0_0+3{\sf S}^0_1-{\sf C}^0_0-{\sf C}^0_1\right)^2
+\frac{11}{6}\left({\sf S}^1_0+{\sf C}^1_0+{\sf S}^1_1+{\sf C}^1_1\right)^2
\nn\\
&+&\frac{38}{15}\left({\sf T}^{11}_0+{\sf T}^{11}_1\right)^2
+\frac{54}{5}\left({\sf T}^{31}_0+{\sf T}^{31}_1\right)^2,
\label{N.11}\er
for the decay $\Lambda n\rightarrow nn$.

\subsubsection{Parity violating contributions}

The PV matrix elements \rf{A5} are reduced to the nuclear moments
\br
{\sf P}_M^{l{\sf l}}(p,P)&=&\A_M{(pl|f^{(-)}_M|1{\sf l})}(P0|10),
\nn\\
{\sf Q}_M^{l{\sf l}}(p,P)&=&\A'_M{(pl|f^{(+)}_M|1{\sf l})}(P0|10),
\label{N.12}\er where the radial integrals $(pl|f^{(\pm)}_M|1{\sf
l})$ are defined in \rf{A8}. Using the notation, \br
\widetilde{{\sf P}}_{\eta}&=&{\sf P}_{\eta}-{\sf Q}_{K_0^*},~~
\widetilde{{\sf P}}_{K_0}={\sf P}_{K_0}-{\sf Q}_{\omega},~~
\widetilde{{\sf P}}_{K_0^*}={\sf P}_{K_0^*}+{\sf P}_{\omega},
\nn\\
\widetilde{{\sf P}}_\pi&=&{\sf P}_\pi-{\sf Q}_{K_1^*},~~
\widetilde{{\sf P}}_{K_1}={\sf P}_{K_1}-{\sf Q}_{\rho},~~
\widetilde{{\sf P}}_{K_1^*}={\sf P}_{K_1^*}+{\sf P}_{\rho},
\label{N.13}\er
we obtain:
\br
{\cal I}_p
&=&2\left(1+R\right)
\left[3\left(\widetilde{{\sf P}}^{10}_\pi\right)^2+\left(\widetilde{{\sf P}}^{10}_\eta\right)^2
+3\left(\widetilde{{\sf P}}^{10}_{K_1}\right)^2
+\left(\widetilde{{\sf P}}^{10}_{K_0}\right)^2
+10\left(\widetilde{{\sf P}}^{10}_{K_1^*}\right)^2+2\left(\widetilde{{\sf P}}^{10}_{K_0^*}\right)^2
\right.
\nn\\
&-&\left.2\widetilde{{\sf P}}^{10}_\eta\widetilde{{\sf P}}^{10}_{K_1}
+2\widetilde{{\sf P}}^{10}_\pi
\left(2\widetilde{{\sf P}}^{10}_{K_1}-\widetilde{{\sf P}}^{10}_{K_0}
+4\widetilde{{\sf P}}^{10}_{K_1^*}-2\widetilde{{\sf P}}^{10}_{K_0^*}\right)
+4\widetilde{{\sf P}}^{10}_{K_1}(2\widetilde{{\sf P}}^{10}_{K_1^*}
-\widetilde{{\sf P}}^{10}_{K_0^*} )
-4\widetilde{{\sf P}}^{10}_{K_1^*}
\widetilde{{\sf P}}^{10}_{K_0^*}
 \right]
\nn\\
&+&14\left(\widetilde{{\sf P}}^{21}_\pi\right)^2+2\left(\widetilde{{\sf P}}^{21}_\eta\right)^2
+8\left(\widetilde{{\sf P}}^{21}_{K_1}\right)^2
+\frac{4}{3}\left(\widetilde{{\sf P}}^{21}_{K_0}\right)^2
+14\left(\widetilde{{\sf P}}^{21}_{K_1^*}\right)^2+\frac{10}{3}\left(\widetilde{{\sf P}}^{21}_{K_0^*}\right)^2
\nn\\
&+&4\widetilde{{\sf P}}^{21}_\eta\widetilde{{\sf P}}^{21}_{K_1}
-4\widetilde{{\sf P}}^{21}_\pi
\left(2\widetilde{{\sf P}}^{21}_{K_0^*}+2\widetilde{{\sf P}}^{21}_{K_1}
-\widetilde{{\sf P}}^{21}_{K_0}
+4\widetilde{{\sf P}}^{21}_{K_1^*}-2\widetilde{{\sf P}}^{21}_{K_0^*}\right)
+4\widetilde{{\sf P}}^{21}_{K_1}\left(\widetilde{{\sf P}}^{21}_\eta-\widetilde{{\sf P}}^{21}_{K_1^*}
+\widetilde{{\sf P}}^{21}_{K_0^*} \right)
\nn\\
&+&4\widetilde{{\sf P}}^{21}_{K_1^*}\left(2\widetilde{{\sf P}}^{21}_\eta
-\widetilde{{\sf P}}^{21}_{K_0^*}\right)
+\frac{4}{3}\widetilde{{\sf P}}^{21}_{K_0^*}\widetilde{{\sf P}}^{21}_{K_0}
\nn\\
&+&\frac{2}{3}\left(\widetilde{{\sf P}}^{01}_{K_0}\right)^2
+6\left(\widetilde{{\sf P}}^{01}_{K_1}\right)^2
+\frac{2}{3}\left(\widetilde{{\sf P}}^{01}_{K_0^*}\right)^2
+6\left(\widetilde{{\sf P}}^{01}_{K_1^*}\right)^2
-4\widetilde{{\sf P}}^{01}_{K_0}\widetilde{{\sf P}}^{01}_{K_1}\nn\\
&-&\frac{4}{3}\widetilde{{\sf P}}^{01}_{K_0^*}
\left(\widetilde{{\sf P}}^{01}_{K_0}-3\widetilde{{\sf P}}^{01}_{K_1}
+3\widetilde{{\sf P}}^{01}_{K_1^*}\right)
+4\widetilde{{\sf P}}^{01}_{K_1^*}
\left(\widetilde{{\sf P}}^{01}_{K_0}-3\widetilde{{\sf P}}^{01}_{K_1}\right),\nn\\
\label{N.14}\er
for the $\Lambda p\rightarrow np$ decay, and
\br
{\cal I}_n&=&\left(3+\frac{43R}{18}\right)
\left[\left(\widetilde{{\sf P}}_\pi^{10}+\widetilde{{\sf P}}_\eta^{10}\right)^2
+\left(\widetilde{{\sf P}}_{K_0}^{10}+\widetilde{{\sf P}}_{K_1}^{10}\right)^2\right]
+\left(4+\frac{14R}{3}\right)
\left(\widetilde{{\sf P}}_{K_0^*}^{10}+\widetilde{{\sf P}}_{K_1^*}^{10}\right)^2
\nn\\
&-&\left(2+\frac{R}{9}\right)\left(
\widetilde{{\sf P}}_\pi^{10}+\widetilde{{\sf P}}_\eta^{10}\right)
\left(\widetilde{{\sf P}}_{K_0}^{10}+\widetilde{{\sf P}}_{K_1}^{10}\right)
\nn\\
&-&\left(4+\frac{14R}{3}\right)
\left(\widetilde{{\sf P}}_{K_0^*}^{10}+\widetilde{{\sf P}}_{K_1^*}^{10}\right)
\left(
\widetilde{{\sf P}}_\pi^{10}+\widetilde{{\sf P}}_\eta^{10}
+\widetilde{{\sf P}}_{K_0}^{10}+\widetilde{{\sf P}}_{K_1}^{10}\right)
\nn\\
&+&\fot\left(\widetilde{{\sf P}}_\pi^{21}+\widetilde{{\sf P}}_\eta^{21}+
\widetilde{{\sf P}}_{K_0}^{21}+\widetilde{{\sf P}}_{K_1}^{21}\right)^2
\nn\\
&+&2\left(\widetilde{{\sf P}}_{K_0^*}^{21}+\widetilde{{\sf P}}_{K_1^*}^{21}\right)
\left(\widetilde{{\sf P}}_{K_0^*}^{21}+\widetilde{{\sf P}}_{K_1^*}^{21}
+\widetilde{{\sf P}}_\pi^{21}+\widetilde{{\sf P}}_\eta^{21}+
\widetilde{{\sf P}}_{K_0}^{21}+\widetilde{{\sf P}}_{K_1}^{21}\right),
\label{N.15}\er
for the $\Lambda n\rightarrow nn$  decay.
\newpage
\section{Numerical Results and Discussion}

The numerical values of the parameters,  defined  in Table \ref{table1}
and   necessary to specify the transition potential, are summarized
in Table \ref{table3}. For the sake of comparison all cutoffs  appearing in \rf{3.17},
as well as all coupling constants,
were taken  from Ref. \cite{Pa97}, where, in turn, the strong couplings
have been taken from Refs. \cite{Na77,Ma89} and the  weak ones from Ref. \cite{Du96}.
The energy difference $\Delta_{j_Nm_{ t_N}}$ in \rf{2.20} is evaluated from the experimental
single nucleon and hyperon energies, quoted in Ref. \cite{Ra92}.

\begin{table}[h]
\caption{Parameters used in the calculations: masses  (in MeV), cut-offs (in GeV) and
the  isoscalar ($\tau=0$) and isovector ($\tau=1$) coupling constants  (in units of
$10^{-11}$ MeV$^{-2}$). \label{table3}}
\begin{tabular}{|c|ccrrrr|}
M&$m_M$&$\Lambda$&$\A_M/m_M$&$\A'_M/m_M$&$\B_M$&$\B'_M/m_M^2$\\
\hline
\hline
&&    &    $\tau=0$ &  &     &         \\
\hline
$\eta$&$548.6 $&$1.3$&$0.247$&$   $&$-0.525$&$  $
\\
&   & &     &       &   &      \\
$K$ &           $495.8$&$ 1.2$&$-0.828$&$   $&$0.228$            &$$
            \\
&    &&     &       &    &     \\
$\omega$&$783.4 $&$1.5$&$-0.274$&$-0.420$&$-0.923$& $-1.395$
\\
& &   &     &       &     &    \\
$K^*$&$892.4$&$ 2.2$&$0.376$&$0.237$&$0.632$&$1.016$
\\
& &   &     &       &      &   \\
\hline
&&    &    $\tau=1$ &       &&         \\
\hline
$\pi$&$140.0$&$ 1.3$&$1.175$&$  $&$-0.546$&$  $
            \\
&  &  &     &       &        & \\
$K$&$495.8$&$ 1.2$&$-0.127$&$   $&$0.764$           &$  $
           \\
&    &&     &       &         &\\
$\rho$&$775.0 $&$1.4$&$0.273$&$0.105$&$-0.907$& $-0.407$
\\
&&    &     &       &        & \\
$K^*$&$892.4$&$ 2.2$&$0.514$&$0.324$&$1.072$&$0.274$
\\
&&    &     &       &         &\\
\end{tabular}
\end{table}

The finite nucleon size (FNS) effects at the interaction vertices are gauged
 by the monopole form factor  $F_M^{\rm(FNS)}(\qb^2)= \displaystyle
\frac{\Lambda_M^2-m_M^2}{\Lambda_M^2+\qb^2}$, which implies  that
the propagators  in \rf{3.10} and \rf{3.11} must be replaced by
\br
f_M(r)\go\overline{f}_M(r)&=&f_M(r)-f_{\Lambda_M}(r)
-\frac{r(\Lambda_M^2- m_M^2)}{2\Lambda_M}f_{\Lambda_M}(r),
\nn\\
f^S_M(r)\go \overline{f}^S_M(r)&=&f^S_M(r)-f^S_{\Lambda_M}(r)
-\frac{1}{6}(\Lambda_M^2- m_M^2)(\Lambda_Mr-2)f_{\Lambda_M}(r),
\nn\\
f^T_M(r)\go \overline{f}^T_M(r)&=&f^T_M(r)-f^T_{\Lambda_M}(r)
-\frac{1}{6}(\Lambda_M^2- m_M^2)(\Lambda_Mr+1)f_{\Lambda_M}(r),
\nn\\
f'_M(r)\go \overline{f}'_M(r)&=&f'_M(r)-f'_{\Lambda_M}(r)
+\frac{r(\Lambda_M^2- m_M^2)}{2}f_{\Lambda_M}(r),
\label{3.17}\er
where $f_{\Lambda_M}(r)$ has the same structure as ${f}_M(r)$ but with $m_M\go
\Lambda_M$.

  The initial and final short range correlations (SRC) are taken into
account, respectively, via the correlation functions \cite{Pa97},
\br
g_i(r)&=&\left( 1 -  e^{-r^2 / \alpha^2} \right)^2 + \beta r^2
e^{-r^2 / \gamma^2},
\nn\\
g_f(r)&=&1-j_0(q_c r),
\label{3.18}\er
with $\alpha=0.5$ fm, $\beta=0.25$ fm$^{-2}$, and  $\gamma= 1.28$ fm,
and $q_c=3.93$ fm$^{-1}$.

It is a general belief nowadays that, in any realistic evaluation of the
 hypernuclear NM  decay, the FNS and SRC have to be included simultaneously.
 Therefore, in the present paper we will discuss  only
 the numerical results, in which both of  these
 renormalization effects are considered.
Under these circumstances, and because of the relative smallness of pion mass,
 the transition is dominated by the OPE  \cite{Pa97}.

\begin{table}
\begin{center}
\caption{Parity conserving (PC) and parity violating (PV)
nonmesonic decay rates for $^{12}_\Lambda C$, in units of
$\Gamma^0= 2.50 \cdot 10^{-6}$ eV. All  coupling
constants and the cutoff parameters are  from Table \ref{table3} and $b=1.51$ fm.
All calculations were done within the EPHM, except for a few results
which were evaluated in the QTDA and are shown parenthetically.
\label{table4}}
\begin{tabular}{|c|cccc|}
Mesons&$\Gamma_n^{\ss\rm PC}$&$\Gamma_n^{\ss\rm PV}$
&$\Gamma_p^{\ss\rm PC}$&$\Gamma_p ^{\ss\rm PV}$\\ \hline
$\pi$&$0.009$ &$0.151$&$0.734$&$0.383$ \\
     &($0.016$) &($   0.153$)&    ($0.732$)&  ($0.373$) \\
$\eta$&$0.003$&$ 0.004 $&$  0.006  $&$  0.003   $\\
$K$&$ 0.008 $&$   0.069  $&$  0.097  $&$  0.043 $\\
$\rho$&$ 0.005  $&$   0.003  $&$   0.109 $&$    0.008 $\\
$\omega$&$  0.004 $&$ 0.007  $&$ 0.066$&$ 0.004$\\
$K^*$&$ 0.025 $&$  0.034$&$     0.056 $&$    0.028$\\
\hline
$\pi+\eta$&$     0.013 $&$     0.204 $&$     0.630 $&$ 0.383 $\\
$\pi+K$&$0.013 $&$    0.258$&$     0.325 $&$    0.512 $\\
$\pi+\rho$&$ 0.009  $&$   0.133 $&$    0.583 $&$    0.461 $\\
$\pi+\omega$&$   0.015 $&$    0.176  $&$   0.902  $&$   0.406 $\\
$\pi+K^*$&$ 0.044  $&$   0.075  $&$   1.020  $&$   0.455  $\\
$\pi+\eta+K$&$0.008  $&$   0.318 $&$    0.259$&$     0.505$\\
            &($0.011  $)&($   0.330 $)&($    0.258$)&($0.516$)\\
$\pi+\eta+K+K^*$&$   0.052 $&$    0.268 $&$    0.486 $&$    0.602$\\
all~mesons&$  0.037 $&$     0.240 $&$     0.347$&$      0.714$\\
&($  0.039 $)&($     0.250 $)&($     0.346$)&($      0.702$)\\
\end{tabular}
\end{center}
\end{table}

The major part of the numerical calculations were done in the EPHM
where the only  free parameter  is  the harmonic oscillator length
$b$. The most commonly used estimate is  $ b= A^{1/6}$ fm
\cite{Boh69,Ri80}, which  corresponds to the oscillator energy
$\hbar\omega=41A^{-1/3}$ MeV, and gives $ b=1.51$ fm. For light
nuclei it is sometimes preferred  to employ
$\hbar\omega=45A^{-1/3}-25A^{-2/3}$ MeV, which yields $ b=1.70$
fm. Moreover, a $\Lambda$ particle in a hypernucleus is typically
less bound than the corresponding nucleon and hence $b_\Lambda$
could be larger than $b_N$. For instance, in Ref. \cite{Pa97} was
used $b=(b_\Lambda+b_N)/2=1.75 $ fm, which comes from $b_N=1.64$
fm and $b_\Lambda=1.87$ fm. As there is no deep  motivation for
 preferring  one particular value of $b$, the numerical results will be exhibited for both
 $b=1.51$ fm and $b=1.75$ fm.

First, a few illustrative  results, obtained in the EPHM (Eq. \rf{N.4}) and
 the simplified version of the QTDA (Eq. \rf{M.3}), are displayed in the Table \ref{table4}.
The  hyperon-nucleon interaction in  the later approach was  taken to be
a simple $\delta$ force, which has been recently used with success as the
nucleon-nucleon interaction to explain the weak decay
processes in $^{12}C$ \cite{Krm02}. The resulting pairing BCS factors were:
$v_{s_{1/2}}=0.9868$, $v_{p_{3/2}}=0.8978$ and $v_{p_{1/2}}=0.6439$.
Although we have expected to obtain
small differences between the EPHM and the QTDA, it came as a surprise
that they turned out to be so tiny. Thus,
 henceforth only the first one will be used.

Next, we  combine  results from Table \ref{table4} with
 the multipole expansion done in the previous section to find out
 the roles played by different mesons. Note that the formulas \rf{N.14} and \rf{N.15}
 depend on the ratio \rf{N.6}, and   it was found numerically that  the approximation
\be
R=1,
\label{R.1}\ee
reproduces  fairly well the exact calculations. This estimate helped us
to formulate   the following comments:

{\bf PC potential};
The  dominant contributions to $\Gamma_p$ and $\Gamma_n$
come from the ${\sf l}=0$ matrix elements, while
the  ${\sf l}=1$ wave
contributes relatively  little: $\cong 2\%$ to $\Gamma_p$
and  $\cong 10\%$ to $\Gamma_n$.
On the other hand, for the parametrization displayed in Table \ref{table3}, one finds that:
(1) the $\omega$ and $K^*$ mesons mainly cancel out in ${\sf C}^0_0$, as   the
 $\rho$ and $K^*$ mesons  do in ${\sf C}^0_1$, and
(2) the matrix elements ${\sf S}^0_0$ and ${\sf S}^0_1$ are  small
in comparison with $(3{\sf T}^{20}_1-{\sf T}^{20}_0)$, which makes
$\Gamma_p$ to be large  vis-$\grave{a}$-vis $\Gamma_n$. Thus,
using the estimate \rf{R.1}, one ends up with the following
approximate result for the PC contributions: \br {\cal I}_p+{\cal
I}_n&\cong&{\cal I}_p\cong 2\I_p^{s_{1/2}}\cong\ 24\left(3{\sf
T}^{20}_1-{\sf T}^{20}_0\right)^2. \label{R.2}\er
From Table
\ref{table3} and \rf{N.9} one can also see that: (i)  the $\omega$
and $K^*$ mesons contribute coherently with the pion,  while the
remaining three mesons contribute out of phase, and (ii) the
different vector meson contributions have the tendency to cancel
among themselves. As is shown in Table \ref{table4}, the overall
effect is a reduction of the pion transition rate by approximately
a factor of two.


{\bf PV potential};
As in the PC case, the dominant PV transition strengths come from the
${\sf l}=0$ wave, through the $\widetilde{{\sf P}}^{10}_M$ moments.
The ${\sf l}=1$ wave  from  the $p_{3/2}$ state gives rise to
 $l=0$ and  $l=2$ outgoing   channels.  The first one can always be neglected,
while
the second one contributes with $15\%$ to $\Gamma_p$ and
with  $2\%$ to $\Gamma_n$, when only the $\pi$ meson is considered.
After including  all mesons  these percentages drop to  $6\%$ and $1\%$,
 respectively. Also here  the partial   $s_{1/2}$ and $p_{3/2}$
  contributions are  approximately  equal for all mesons in the proton induced channel
and notably different in the neutron induced channel.

From  Table \ref{table4} it is  easily found that the most
important PV contributions arise from the
 ${\sf P}^{10}_\pi$ moment, and from its interference with the
 ${\sf P}^{10}_\eta$, ${\sf P}^{10}_K$ and ${\sf P}^{10}_{K^*}$ moments. Thus,
retaining only the most relevant terms in \rf{N.14} and \rf{N.15},
the following rough estimates are obtained:
\br
{\cal I}_p\cong 2\I_p^{s_{1/2}}
&\cong& 4\left[3\left({\sf P}^{10}_\pi\right)^2
+3\left({\sf P}^{10}_{K_1}\right)^2+\left({\sf P}^{10}_{K_0}\right)^2\right]
\nn\\&&
+8{\sf P}^{10}_\pi\left(2{\sf P}^{10}_{K_1}-{\sf P}^{10}_{K_0}
+4{\sf P}^{10}_{K_1^*}-2{\sf P}^{10}_{K_0^*}
-3{\sf Q}^{10}_{K_1^*}\right),
\label{R.3}\er
and
\br
{\cal I}_n
&\cong&\frac{97}{18}\left[\left({\sf P}^{10}_\pi\right)^2
+\left({\sf P}_{K_1}^{10}+{\sf P}_{K_0}^{10}\right)^2\right]
-\frac{19}{9}{\sf P}^{10}_\pi\left({\sf P}_{K_1}^{10}+{\sf P}_{K_0}^{10}\right)
\nn\\&&
+\frac{97}{9}{\sf P}^{10}_\pi
\left({\sf P}_\eta^{10}-{\sf Q}_{K^*_1}^{10}-{\sf Q}_{K^*_0}^{10}\right)
-\frac{78}{9}{\sf P}^{10}_\pi\left({\sf P}_{K^*_1}^{10}+{\sf P}_{K^*_0}^{10}\right).
\label{R.4}\er
 These relations are notably more complicated than  \rf{R.2}.
Nevertheless, it can be concluded  that: (1)
the $\eta$ meson is only significant for $\Gamma_n$, and (2)
the $K$ and $K^{*}$
mesons increase both transition rates, but in a different way.

\begin{figure}[h]
\begin{center}
\hspace{1.7cm}
    \leavevmode
   \epsfxsize = 8cm
   \epsfysize = 9cm
  \epsffile{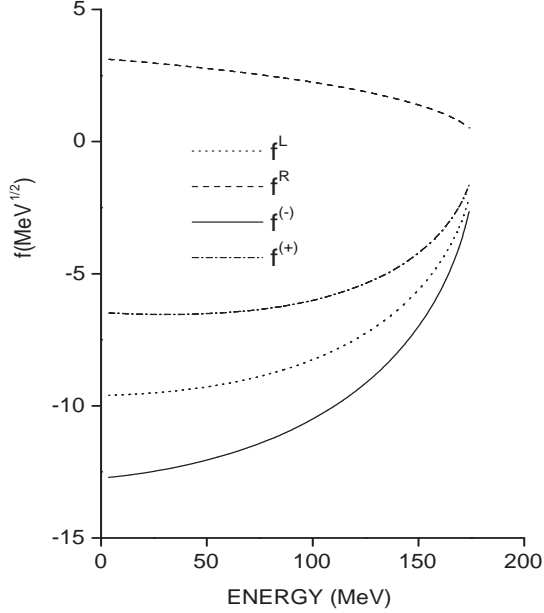}
   \end{center}
 {\tighten \caption{Matrix elements of the radial operators $f^L_{K^*}$, $f^R_{K^*}$,
 $f^{(+)}_{K^*}$, and $f^{(-)}_{K^*}$, as a function of the energy.
 }} \label{fig2}
\end{figure}
Before proceeding it is worth to say a few
words on the "new" nuclear moments ${\sf Q}_M^{10}$ and compare them with
the well known moments ${\sf P}_M^{10}$. As seen from \rf{N.12} and \rf{A8}-\rf{A10}
 they basically differ in the radial dependence.
Specifically, we discuss the radial matrix element
\be
\overline{(p,1|f^{(+)}_{K^*}|10)}=\overline{(p,1|f^{L}_{K^*}|10)}+
\overline{(p,1|f^{R}_{K^*}|10)},
\label{R.5}\ee
which appears in  ${\sf Q}_{K^*}^{10}$, together with the usual matrix element
\be
\overline{(p,1|f^{(-)}_{K^*}|10)}=\overline{(p,1|f^{L}_{K^*}|10)}-
\overline{(p,1|f^{R}_{K^*}|10)},
\label{R.6}\ee
which is contained in ${\sf P}_{K^*}^{10}$.
The overline indicates that both the FNS and  SRC are included, as explained in the
Appendix.

As can be seen from  Fig. 2, the matrix elements of $f^{L}_{K^*}$ and $f^{R}_{K^*}$ have
opposite signs, and as a consequence the matrix element of  $f^{(-)}_{K^*}$ is  larger
in magnitude  than that of $f^{(+)}_{K^*}$.
 A   rough approximation for the mean values is:
\br
\left|\left<\overline{(p1|f^{(+)}_{K^*}|10)}\right>\right|
\cong \fot\left|\left<\overline{(p1|f^{(-)}_{K^*}|10)}\right>\right|.
\label{R.7}\er
As  $\A_{K^*}\gsim\A'_{K^*}$ (see Table \ref{table3}) we end up with the
estimate:
\br
\left|\left<{\sf Q}^{10}_{K^*}\right>\right|
\cong 0.3\left|\left<{\sf P}^{10}_{K^*}\right>\right|.
\label{R.8}\er
Thus, equations \rf{R.3} and \rf{R.4} show that the $K^*$ meson mainly contributes through
the moments ${\sf P}^{10}_{K^*}$, augmenting the magnitude of
$\Gamma_p^{\ss\rm PV}$ and diminishing  that of $\Gamma_n^{\ss\rm PV}$.
 The matrix elements ${\sf Q}^{10}_{K^*}$, in contrast, reduce both
 transition rates.

Furthermore, the equation   \rf{N.13} indicates that each  vector  moment
${\sf Q}^{10}_{\rho,\omega,K^*}$   is accompanied by a pseudo-scalar  moment
${\sf P}^{10}_{\pi,\eta,K}$. Both  integrals $(p1|f_M^{(\pm)}|10)$ are
negative for all mesons. Then, using the values of the coupling constants $\A_{\pi,\eta,K}$ and
$\A'_{\rho,\omega,K^*}$ listed in Table \ref{table3},
it can be inferred  that ${\sf Q}^{10}_{\rho,\omega,K^*}$ and
${\sf P}^{10}_{\pi,\eta,K}$ moments mostly add incoherently.

\begin{table}[t]
\begin{center}
\caption{Parity conserving (PC) and parity violating (PV)
nonmesonic decay rates for $^{12}_\Lambda C$, in units of
$\Gamma^0= 2.50 \cdot 10^{-6}$ eV. The data are taken from Refs.
\protect\cite{Mo74,Sz91,No95,Bh98,Ha02}, and
large experimental errors are due to the low
efficiencies and large backgrounds in neutron detection.
The calculations were performed for both  $b=1.51$ fm and  $b=1.75$ fm, being the later
given parenthetically. In {\it Calculation A} all parameters are  from Table \ref{table3},
and   PS and V
stand, respectively, for the pseudo-scalar ($\pi+\eta+K$) and the
vector ($\rho+\omega+K^*$) mesons, while  the label  $ ({\sf P})$  indicates
that only the moments ${\sf P}_M$ are considered
(see \rf{N.12}). In
{\it Calculation B} the coupling constants listed in Table \ref{table3}  are modified as:
 $\A_\eta\go 3\A_\eta$, $\A_{K_1}\go 5\A_{K_1}$,
 $\B_\eta\go 3\B_\eta$ and $\B_{K}\go 2\B_{K}$, and the signs of all vector meson potentials
 are inverted.
\label{table5}}
\begin{tabular}{|c|ccc|}
&$\Gamma_{NM}=\Gamma_n+\Gamma_p$&$\Gamma_p$&$\Gamma_{n/p}=\Gamma_n/\Gamma_p$\\
\hline
$Measurements$:&&&\\
Ref. \cite{Mo74}&&&$0.70\pm 0.3$\\
Ref. \cite{Mo74}&&&$0.52\pm 0.16$\\
Ref. \cite{Sz91}&$1.14\pm 0.2$&&$1.33^{+1.12}_{-0.81}$\\
Ref.\cite{No95}& $0.89\pm 0.15\pm 0.03$& $0.31^{+0.18}_{-0.11}$ & $1.87\pm 0.59^{+0.32}_{-1.00}$ \\
Ref. \cite{Bh98}&$1.14\pm 0.08$& &\\
Ref. \cite{Ha02}&&&$1.17^{+0.09+0.22}_{-0.08-0.18}$\\
\hline
$Calculation~A$:&&&\\
$\pi $&$1.277(1.006)$&$1.116(0.885)$&$ 0.143(0.137)$ \\
$PS $&$1.100(0.851)$&$0.774(0.601)$&$ 0.420(0.416) $\\
$PS+K^*$&$1.408(1.091) $&$1.088(0.846)$&$0.294(0.290) $\\
$PS+V$&$  1.338(1.038) $&$1.061(0.825)$&$     0.261(0.259)$\\
$PS+V ({\sf P})$&$1.539(1.190) $&$1.196(0.927)$&$     0.287(0.284)$\\
\hline $Calculation~B$:&&&\\
$PS' $&$     1.145(0.874)$&$ 0.555(0.419)$&$  1.064(1.089)$\\
$PS'-K^*$&$1.273(0.971)$&$ 0.540(0.407)     $&$     1.355(1.384) $\\
$PS'-V$&$    1.297(0.989)$&$0.542(0.408) $ &$    1.394(1.423)$\\
\end{tabular}
\end{center}
\end{table}


The experimental results  for the total transition  rate $\Gamma_{NM}$, the proton
partial width $\Gamma_{p}$, and the ratio
$\Gamma_{n/p}$  in  $^{12}_\Lambda C$ are displayed in Table \ref{table5}.
In the same table the theoretical estimates  are also shown,  grouped as:
\bnu
\item {\it Calculation A}. All the  parametrization
is taken from   Table \ref{table3}, and the following cases are shown and commented:
\bit
\item $(\pi)$:
The simple OPE model accounts  for $\Gamma_{NM}$, but it badly fails regarding
$\Gamma_{p}$ and $\Gamma_{n/p}$.
\item $(PS)$:
When   $\eta$ and $K$ mesons are included, the
total transition rate is  only slightly modified, while  $\Gamma_{p}$
and $\Gamma_{n/p}$
change significantly, coming somewhat closer to the measured values.
\item $(PS+K^*)$:
The  incorporation of the $K^*$ meson increases  $\Gamma_{NM}$ and
  $\Gamma_{p}$, decreases $\Gamma_{n/p}$,
 and in this way   worsens  the agreement with the data.
\item
$(PS+V)$: The results are not drastically modified when all vector mesons are built-in.
\item $(PS+V({\sf P}))$:
 All 6 mesons are included, but only the PV moments ${\sf P}_M$ are considered.
The importance of the new moments ${\sf Q}_M$ is evident from the comparison with
the previous case.
\eit
The main conclusion is that it is not possible  to reproduce
simultaneously  the data for all three observables
 $\Gamma_{NM}$,  $\Gamma_{p}$ and $\Gamma_{n/p}$, when the BBM
 coupling are constrained by the $SU(3)$  and $SU(6)_w$ symmetries.
\item {\it Calculation B}.
We discuss now what happens when the just mentioned  constraints are relaxed,
and  the FNS and SRC parametrizations,   as well as the
  the pion couplings, are kept unchanging. That is, the transition
  potential is considered to  be given by a series of Yukawa like potentials
  with different spin and isospin dependence.
  The simple increase of the $K$ coupling does not solve the
  problem by itself.
For instance,  for $\A_{K}\go 2\A_{K}$ and  $\B_{K}\go 2\B_{K}$,
the contribution of all three pseudo-scalar  mesons is (when $b=1.51$ fm):
$\Gamma_{NM}=1.404~\Gamma^{0}$,
$\Gamma_{p}=0.815~\Gamma^{0}$ and $\Gamma_{n/p}=0.723$, and when the vector
 mesons are added one gets:
$\Gamma_{NM}=1.714~\Gamma^{0}$,
$\Gamma_{p}=1.130~\Gamma^{0}$ and $\Gamma_{n/p}=0.518$. Namely, $\Gamma_{NM}$
turns out to be too large.
But from the previous  discussion, in relation to equations \rf{R.2}, \rf{R.3}
 and \rf{R.4}, we have learned that it could be possible to reproduce at the same time
 the data for all three observables
 by: (i) making  the total tensor interaction in  $\Gamma_p^{\ss\rm PC}$ small, and
 simultaneously
 (ii)  decreasing $\Gamma_p^{\ss\rm PV}$ and increasing $\Gamma_n^{\ss\rm PV}$, without
modifying  $\Gamma_{\ss\rm NM}$ too much.
The first goal can be accomplished, for instance, through
 the  modifications:
 $\B_\eta\go 3\B_\eta$ and $\B_{K}\go 2\B_{K}$, and the second one with
 $\A_\eta\go 3\A_\eta$ and $\A_{K_1}\go 5\A_{K_1}$.
 The following cases are illustrated in Table \ref{table5}:
\bit
\item $(PS')$: Only the pseudoscalar mesons are included with the above
changes in  $\eta$ and $K$ meson couplings.
\item $(PS'-K^*)$: The $K^*$ meson potential  is incorporated but with the inverted  sign.
 \item $(PS'-V)$: All vector meson potentials are included  with the inverted signs.
\eit
\enu
No best fit to data has been attempted. Yet, it is clear that there are many other
set of parameters that reproduce reasonable well the data.
We wish  to stress as well that,
when the vector mesons are considered, the correct values of $\Gamma_{n/p}$ are obtained
only by overturning  the signs of the vector meson potentials.

\newpage
\section{Summary and Conclusions}

A novel shell model formalism for the nonmesonic weak decay of the
hypernuclei has been developed.
It involves a partial wave expansion of the emitted nucleon waves
and  preserves   naturally  the antisymmetrization between the
escaping particles and the residual  core. The general expression \rf{2.13} is valid
for any nuclear model and it shows that the NM transition rates should depend,
in principle, on both: (i) the weak transition potential, through the
 elementary transition amplitudes $\M(pPlL\lambda SJT;{j_\Lambda j_N,m_{t_N}})$,
 and (ii) the nuclear structure,
through the two-particle $N\Lambda$ parentage coefficients
$\Bra{J_I}\left( a_{j_N m_{ t_N}}^\dag a_{j_\Lambda }^\dag\right)_{J}\Ket{J_F}$.
 The explicit evaluation of the matrix elements $\M$ is illustrated as well.

Two nuclear models for  even-mass hypernuclei, namely  the EPHM and the QTDA,
were  worked out in detail, and the Eqs. \rf{N.4} and
 \rf{M.2} were derived.
 The last one explicitly depends on the initial and final
 wave functions. But, because of: i) the inclusive nature of the nonmesonic decay, and
 ii) the peculiar properties of the coefficients $F_{m_{t_b}J}^{j_b}(j_a)$
this   dependence is totally washed out for all practical purposes.
In this way we have arrived at a very simple  result for
transition rates, given by the Eq. \rf{M.4}, which except for the BCS pairing factors
$v^{2}_{j_b}$, agrees with the EPHM result.
Thus, it can be stated that the two-particle correlations in the initial and final state
are only of minor importance if of any.
With some additional effort can also  be incorporated the higher order nuclear structure
effects such as
the four quasi-particle excitations, collective vibrations, rotations,\etc..
 Yet, it is hard to imagine a scenario where the later  could  be relevant at the same time
that  the former are not.
Therefore, we conclude that the  nuclear structure
manifests  basically through the factor $F_{m_{t_b}J}^{j_b}(j_a)$, which is
engendered by the Pauli principle;
$j_a$ stands for the hyperon partner in the initial state, and $j_b$ runs over all
proton and neutron occupied states in the initial nucleus.
It is amazing to notice that the Eq. \rf{M.4} is valid for any even-mass system, which can
be so light as   ${^{4}_\Lambda H}$ and ${^{4}_\Lambda He}$ are or so heavy as
${^{208}_\Lambda Pb}$ is. (A quite similar result is also obtained for the  odd-mass hypernuclei,
and this issue will be discussed elsewhere.) One should also add that the last equation
contains the same physics as the Eq. (5) in Ref. \cite{Pa97} or the Eq. (30) in
Ref. \cite{It02}, with the advantage that we do not have to deal with  spectroscopic
factors. Of course, neither the initial and
final wave functions are needed.

Attention has been given to the nonrelativistic approximation,
used to derive the weak effective
hypernuclear one-meson exchange potentials \rf{3.14} and \rf{3.15}.
Same errors and misprints that appear in
the recent papers \cite{Pa97,Pa99,Pa01} have been corrected.   Additional parity
violating vector meson  operators $\mbs_N\cdot\fb^{(+)}(r)$ and  $\mbs_\Lambda\cdot\fb^{(+)}(r)$,
usually neglected, have been considered as well.
The  matrix elements of these new terms were fully
discussed, and it was found  that they are quite important quantitatively
and therefore should not be omitted.

With the OMEM parametrization from the literature \cite{Pa97},
and keeping the treatment of the FSI at the simple Jastrow like level ($g(r)=1-j_0(q_c r)$),
 we  reproduce satisfactorily  the data for  the total
transition rate ($\Gamma_{NM}^{\rm th}\cong \Gamma^0$), but the $n/p$-ratio
($\Gamma_{n/p}^{\rm th}\lsim 0.42$) and the proton partial width
($\Gamma_p^{\rm th}\gsim 0.60 \Gamma^0$) are not well accounted for.
More elaborate treatments of the FSI  increase  sensibly the $n/p$-ratio,
but they are unable to solve the puzzle \cite{Pa97,Pa01}, especially
after the last experimental result for this observable \cite{Ha02}.
We have found that the new vector meson  operators are
not of much helpful in this regard  either.

Finally,  bearing in mind the phenomenological nature of
the OMEM, we have also tried to reproduce all three data  simultaneously by
 varying  the coupling strengths in a significant way.
As the only guide were used the simple formulas \rf{R.4}, \rf{R.5} and \rf{R.6},
which come out from the multipole expansion done within the EPHM.
Such an attempt was  successful, and we get:
$0.87\lsim\Gamma^{\rm th}_{NM}/ \Gamma^0\lsim 1.30$, $1.06\lsim\Gamma^{\rm th}_{n/p}\lsim 1.42$, and
$0.41\lsim\Gamma_p^{\rm th}/ \Gamma^0\lsim  0.55$.
We are conscious  that  changing a coupling  by up to a factor of 5, with the sole
justification
of accounting for the data, is  a rather a desperate way out of the $\Gamma_{n/p}$ puzzle.
Thought no profound physical significance is attached to
 the "new" parameters, it  even can be said that such a procedure is not physical. However, after
having acquired full control of the nuclear structure involved in the process,
and after having convinced ourselves that the  nuclear structure  correlations
can not play a crucial role, we firmly believe that the currently used OMEM should be
radically changed. Either  its parametrization has to   be  modified
or  additional degrees of
freedom have to be  incorporated, such as  the correlated $2\pi$ from Ref. \cite{It02}
or the  $4$-barion point interaction from Ref. \cite{Jun02}, avoiding clearly the double
counting. In fact, it would be very
nice to see the outcomes of such studies.

\bigskip
The authors acknowledge  the support of   ANPCyT (Argentina) under
grant BID 1201/OC-AR (PICT 03-04296), of Fundaci\'on Antorchas
(Argentina) under grant  13740/01-111, and of Croatian Ministry of
Science and Technology under grant 119222. D.T. thanks the Abdus Salam ICTP
Visiting Scholar programme for two travel fares. F.K. and C.B. are
fellows of the CONICET  Argentina. One of us (F.K.) would like to thank
A.P. Gale\~ao for very helpful and illuminating discussions, and for a careful and
critical reading of the manuscript.

\newpage
\appendix
\section*{Nuclear Matrix Elements}


Here we evaluate the transition  matrix elements that appear in \rf{2.15} for
the potentials $\V(\rb,s_1s_2)$ defined in \rf{3.14} and \rf{3.15}.

The PC potential contains the operators $f_M(r)$, $f^S_M(r)(\mbs_\Lambda\cdot\mbs_N)$,
  and $f^T_M(r)S_{\Lambda N}(\rh)$, and the corresponding matrix elements read:
\br ({pPlL\lambda SJ}|f_M(r)|{1{\sf lNL} \lambda' S'J})
&=&\delta_{l{\sf
l}}\delta_{\lambda\lambda'}\delta_{SS'}\delta_{L{\sf{L}}}(PL|{\sf{NL}})
(pl|f_M(r)|1{{\sf{l}}}),
\nn\\\nn\\
({pPlL\lambda SJ}|f^S_M(r)(\mbs_\Lambda\cdot\mbs_N)|{1{\sf lNL}
\lambda' S'J}) &=&\delta_{l{\sf
l}}\delta_{\lambda\lambda'}\delta_{SS'}\delta_{L{\sf{L}}}(PL|{\sf{NL}})
(pl|f^S_M(r)|1{{\sf{l}}})[2S(S+1)-3],
\nn\\\nn\\
({pPlL\lambda SJ}|f^T_M(r)S_{\Lambda N}(\rh)|{1{\sf lNL} \lambda'
S'J}) &=&(-)^{L+l
+J+1}\delta_{SS'}\delta_{L{\sf{L}}}\delta_{S1}\sqrt{120}
\hat{\lambda}\hat{\lambda'}\hat{l} (PL|{\sf{NL}})
\nn\\\nn\\
&\x&
(pl|f^T_M(r)|1{{\sf{l}}})\sixj{\lambda'}{1}{J}{1}{\lambda}{2}
\sixj{l}{{\sf l}}{2}{\lambda'}{\lambda}{L}
(l020|{\sf l}0),
\nn\\
\label{A1}\er with \be (PL|{\sf{NL}}) =\delta_{L{\sf{L}}}\int
R^2dRj_L(PR)\R_{\sf{NL}}(R), \label{A2}\ee and \br
(pl|f_M|1{\sf{l}}) &=&\int r^2dr
j_l(pr){f}_M(r)\R_{1\sf{l}}(r),etc.
 \label{A3}\er

The PV potentials  are of the form
 \begin{eqnarray}
&&\V^{PV}(\rb,s_1,s_2)\sim\Sb\cdot\fb^{(\pm)}(r)~~~~\mbox{with}~~~~~
\Sb=
\left\{
\begin{array}{l}
 \mbs_\Lambda \\
   \mbs_N\\
   i\mbs_\Lambda\x \mbs_N
\end{array}\right.,
\label{A4}\er and we obtain: \br ({pPlL\l
SJ}|\V_M^{PV}|{{\sf{nlNL}}\l' S'J'})
&=&\delta_{L{\sf{L}}}\hat{\l}\hat{\l'}\hat{{\sf{l}}}({\sf{l}}010|l0)
\sixj{\sf{l}}{{\sf{L}}}{\l'}{\l}{1}{l}\sixj{\l'}{S'}{J}{S}{\l}{1}
\nn\\\nn\\&\x&(-)^{J+S+l+L}
\Bra{S}\Sb_M\Ket{S'}{(pl|f_M^{(\pm)}|{{\sf{nl}}})}(PL|{\sf{NL}}).
\label{A5}\er The spin dependent matrix elements are: \br \Bra{S}
\mbs_N \Ket{S'}&=&\sqrt{6}\hat{S}\hat{S}'(-)^{S}
\sixj{1/2}{1/2}{S'}{1}{S}{1/2}=(-)^{S+S'}\Bra{S}
\mbs_\Lambda\Ket{S'}, \label{A6}\er and \be \Bra{S}i
(\mbs_\Lambda\x\mbs_N) \Ket{S'}=\sqrt{12}
(\delta_{S0}\delta_{S'1}+\delta_{S1}\delta_{S'0})=-\Bra{S}i
(\mbs_N\x\mbs_\Lambda) \Ket{S'}. \label{A7}\ee The matrix elements
$(pl|f_M^{(\pm)}|{{\sf{nl}}})$ are easily evaluated and one
obtains, \be
(pl|f_M^{(\pm)}|{{\sf{nl}}})=(pl|f^L_M|{{\sf{nl}}})\pm(pl|f^R_M|{{\sf{nl}}}).
\label{A8}\ee with \br (pl|f^R_M|{{\sf{nl}}}) &\equiv&\int r^2dr
j_l(pr)f_M(r)
\left(\frac{1}{r}\frac{d}{dr}r+\frac{{\sf{l}}({\sf{l}}+1)-l(l+1)}{2r}\right)\R_{\sf{nl}}(r),
\label{A9}\er and \br (pl|f^L_M|{{\sf{nl}}}) &\equiv&-\int
r^2dr\R_{\sf{nl}}(r)f_M(r)
\left(\frac{1}{r}\frac{d}{dr}r+\frac{l(l+1)-{\sf{l}}({\sf{l}}+1)}{2r}\right)j_l(pr).
\label{A10}\er Note that  the "sum rule" \be
(pl|f^L_M|{{\sf{nl}}})-(pl|f^R_M|{{\sf{nl}}})=(pl|f'_M|{{\sf{nl}}})
=\int r^2dr j_l(pr)f'_M(r)\R_{\sf{nl}}(r), \label{A11}\ee should
always be obeyed.

The  radial integral \rf{A10} can be expressed as
\br
&&(pl|f^L_M|{{\sf{nl}}})
=-p\int r^2dr\R_{\sf{nl}}(r)f_M(r)
\nn\\\nn\\
&\x&\left\{
\frac{(l+2)(l+1)-{\sf{l}}({\sf{l}}+1)}{2(2l+1)}j_{l-1}(pr)+
\frac{l(l-1)-{\sf{l}}({\sf{l}}+1)}{2(2l+1)}j_{l+1}(pr)\right\},
\label{A12}\er
which immediately leads to
\br
(pl|f^L_M|{{\sf{nl}}})&=&
\mp p\int r^2drj_{\sf l}(pr)f_M(r)\R_{\sf nl}(r)
;\;\mbox{for}\;l={\sf l}\pm 1.
\label{A13}\er

We are interested here only in
\be
\R_{1l}=(\pi b^2)^{-1/4}\sqrt{\frac{l!}{(2l+1)!}}
\left(\frac{2}{b}\right)^{l+1}r^l \exp(-\frac{r^2}{2b^2}),
\label{A14}\ee
and, in order to simplify the  integral  \rf{A9}, the following relationship can be used
\br
\frac{1}{r}\frac{d}{dr}r\R_{1l}
&=&\left(\frac{l+1}{r}-\frac{r}{b^2}\right)\R_{1l}.
\label{A15}\er
We obtain
\br
&&(pl|f^R_M|{1{\sf{l}}})
=-\frac{1}{{\sf b}^2}\left\{\begin{array}{cc}
 \int r^3dr j_{l}(pr)f_M(r)\R_{1{\sf l}}(r);&
 \;\;\mbox{for}\;l={\sf l}+1 \;,\\\\
 \int (r^2-\hat{{\sf l}}^2{\sf b}^2)rdr j_{l}(pr)f_M(r)\R_{1{\sf l}}(r);&
 \;\;\mbox{for}\;l={\sf l}-1 \;.\\
\end{array}\right.
\label{A16}\er

It should be remembered that the radial wave functions $\R_{{\sf nl}}(r) $
and $\R_{{\sf NL}}(R)$ have to be  evaluated
with harmonic oscillator parameters $\sqrt{2}b$
and $b/\sqrt{2}$, respectively, being $b$ the oscillator length
for the harmonic mean field potential.

As indicated in \rf{3.17} the FNS effects are incorporated directly in the radial integrals
through the replacements  $f_M(r)\go\overline{f}_M(r)$,\etc. At variance, the SRC,
given by \rf{3.18}, are added by the substitutions
\be
\ket{{1{\sf{lm}}}}\go
\overline{\ket{{{1\sf{lm}}}}}=g_i(r)\ket{1{\sf{lm}}};~~~
\ket{plm}\go \overline{\ket{plm}}=g_f(r)\ket{plm},
 \label{A17}\ee
in \rf{A1} and \rf{A3}, and when the FNS and the SRC are included simultaneously,
the radial integrals \rf{A3} become:
\br
\overline{(pl|f_M|1{\sf{l}})}
&=&\int r^2dr j_l(pr)g_f(r)
\overline{f}_M(r)g_i(r)\R_{1\sf{l}}(r).
 \label{A18}\er
Thus, it is equivalent to  comprise  the
SRC either  through the  wave functions, as done in \rf{A17},
 or by renormalizing  the radial form factor:
 $\overline{f}_M(r)\go g_f(r) \overline{f}_M(r)g_i(r)$.
The same is valid for  $f'_M(r)$, $f^S_M(r)$ and $f^T_M(r)$.
On the contrary, for  the integrals \rf{A9} and \rf{A10}
which contain derivatives, from \rf{A17} one has:
\br
\overline{(pl|f^L_M|1{{\sf{l}}})}&=&-
\int r^2dr[j_l(pr)g'_f(r)\pm pj_{\sf l}(pr)g_f(r)]\overline{f}_M(r)g_i(r)\R_{1{\sf l}}(r)
;\;\mbox{for}\;l={\sf l}\pm 1,
\nn\\\nn\\
\overline{(pl|f^R_M|{1{\sf{l}}}})
&=&\left\{\begin{array}{cc}
 \int r^2dr j_{l}(pr)g_f(r)\overline{f}_M(r)[g'_i(r)-r{\sf b}^{-2}g_i(r)]\R_{1{\sf l}}(r);&
 \;\;\mbox{for}\;l={\sf l}+1 \;,\\\\

 \int r^2dr j_{l}(pr)g_f(r)
 \overline{f}_M(r)[g'_i(r)-(r{\sf b}^{-2}-\hat{{\sf l}}^2r^{-1})g_i(r)]\R_{1{\sf l}}(r);&
 \;\;\mbox{for}\;l={\sf l}-1 \;,\\
\end{array}\right.
\nn\\
\label{A19}\er
being $g'(r)\equiv dg(r)/dr$. In this case it is no longer possible  to include
the SRC via the form factor,
which is a direct consequence of the
fundamental difference between the FNS effects and the SRC.
Namely, while the SRC modify the nuclear wave functions, the FNS renormalization
is done directly on the vertices of the Feynman diagrams that determine  the
one-meson exchange transition  potential.

Finally, the  isospin matrix elements needed in the calculation are:
\br
\begin{array}{ll}
\bra{1,-1}{\cal T}_0\ket{{ -\fot},{-\fot}}=1;
 & \bra{1,-1}{\cal T}_1\ket{{ -\fot},{ -\fot}}=1, \\
\\
  \bra{0,0}{\cal T}_0\ket{{ -\fot},{ \fot}}=-\frac{1}{\sqrt{2}};
   &\bra{1,0}{\cal T}_0\ket{{ -\fot},{ \fot}}=\frac{1}{\sqrt{2}},  \\
\\
 \bra{0,0}{\cal T}_1\ket{{ -\fot},{ \fot}}=\frac{3}{\sqrt{2}};
 & \bra{1,0}{\cal T}_1\ket{{ -\fot},{ \fot}}=\frac{1}{\sqrt{2}}.
\end{array}
\label{A20}\er

\newpage


\end{document}